\documentclass{article}
\usepackage{graphicx} 

\usepackage[margin=1in]{geometry}
\usepackage{pdflscape}
\usepackage{titlesec}
\usepackage[LGR,T1]{fontenc}
\usepackage{hyperref}
\hypersetup{
    colorlinks,
    linkcolor={red},
    citecolor={orange},
    urlcolor={blue!80!black}
}

\makeatletter
\def\bbordermatrix#1{\begingroup \m@th
  \@tempdima 4.75\p@
  \setbox\z@\vbox{%
    \def\cr{\crcr\noalign{\kern2\p@\global\let\cr\endline}}%
    \ialign{$##$\hfil\kern2\p@\kern\@tempdima&\thinspace\hfil$##$\hfil
      &&\quad\hfil$##$\hfil\crcr
      \omit\strut\hfil\crcr\noalign{\kern-\baselineskip}%
      #1\crcr\omit\strut\cr}}%
  \setbox\tw@\vbox{\unvcopy\z@\global\setbox\@ne\lastbox}%
  \setbox\tw@\hbox{\unhbox\@ne\unskip\global\setbox\@ne\lastbox}%
  \setbox\tw@\hbox{$\kern\wd\@ne\kern-\@tempdima\left[\kern-\wd\@ne
    \global\setbox\@ne\vbox{\box\@ne\kern2\p@}%
    \vcenter{\kern-\ht\@ne\unvbox\z@\kern-\baselineskip}\,\right]$}%
  \null\;\vbox{\kern\ht\@ne\box\tw@}\endgroup}
\makeatother

\usepackage{graphicx}
\usepackage{subcaption}
\usepackage{tikz}
\usepackage{pgfplots}
\pgfplotsset{compat=1.18} 
\usetikzlibrary{angles,
                graphs,
                graphs.standard, 
                quantikz2,
                decorations.pathmorphing,
                patterns, 
                patterns.meta,
                shadows}
\tikzset{snake it/.style={decorate, decoration=snake}}
\usepackage{soul}
\usepackage{physics}
\usepackage{amsmath,amsfonts}
\usepackage{pifont}

\usepackage{csvsimple}
\usepackage{listings}
\usepackage{minted}

\usepackage{paralist}

\usepackage{booktabs}
\usepackage{makecell}

\usepackage{amsthm}
\usepackage{mdframed}
    \newmdtheoremenv{defn}{Definition}
    \newmdtheoremenv{thrm}{Theorem}
    \newmdtheoremenv{thrm*}{Theorem}
    \newmdtheoremenv{hlo}{Overview}
    \newmdtheoremenv{lmma}{Lemma}
    \newmdtheoremenv{prop}{Proposition}
    \newmdtheoremenv{stm}{Statement}
    \newmdtheoremenv{crl}{Corollary}

\usepackage{algorithm}
\usepackage{algpseudocode}   

\DeclareMathAlphabet{\mathgtt}{LGR}{cmtt}{m}{n}

\usepackage[toc,page]{appendix}

\usepackage[
    backend=biber,
    style=nature,
    sorting=none,
    url=false,
    doi=false,
    isbn=false
]{biblatex}

\DeclareSourcemap{
  \maps[datatype=bibtex]{
    \map{
      \step[fieldsource=journalabbrev]
      \step[fieldset=journal, origfieldval]
    }
  }
}
\newcommand\blfootnote[1]{%
  \begingroup
  \renewcommand\thefootnote{}\footnote{#1}%
  \addtocounter{footnote}{-1}%
  \endgroup
}

\usepackage[affil-it]{authblk}

\title{\large{\textbf{Theoretical Guarantees of Variational Quantum Algorithm with Guiding States}}}

\author{
\normalsize Tuyen Nguyen,$^{1,2,\star}$ and
\normalsize M\'aria Kieferov\'a$^{1,2}$\\
\small $^{1}$Centre for Quantum Software and Information, School of Computer Science,\\
\small Faculty of Engineering and Information Technology,\\
\small University of Technology Sydney, Sydney, NSW 2007, Australia\\
\small $^{2}$Centre for Quantum Computation and Communication Technology,\\
\small University of Technology Sydney, Sydney, NSW 2007, Australia
}
\date{}

\begin{document}

\maketitle

\blfootnote{$^\star$ \href{mailto:tuyen.q.nguyen@student.uts.edu.au}{tuyen.q.nguyen@student.uts.edu.au}}

\begin{abstract}

{Machine learning provides a powerful framework for predicting ground-state properties across families of quantum many-body problems, enabling amortized inference and reducing the cost of repeated simulations. Variational quantum algorithms (VQAs) are promising candidates for implementing such learnable solvers on quantum computers, yet rigorous guarantees for convergence and generalization remain scarce. Motivated by the fact that standard quantum phase estimation (QPE) provides provable performance when given a guiding state with non-trivial overlap with the ground state, we introduce a variational quantum algorithm with guiding states aiming towards predicting ground-state properties of quantum many-body systems. We then develop a proof technique—the linearization trick—that maps the training dynamics of the algorithm to those of a kernel model. This connection yields theoretical guarantees on both convergence and generalization for the VQA under the guiding state assumption. Our analysis shows that guiding states facilitate convergence, suppress finite-size error terms, and ensure stability across system dimensions. Finally, we validate our findings with numerical experiments on 2D random Heisenberg models.}

\end{abstract}



\section*{Introduction}

{Machine learning (ML) has recently emerged as a transformative paradigm for accelerating the study of complex quantum many-body systems. Specifically, ML approaches have demonstrated significant promise in estimating ground-state energies, observables, and phase transitions across diverse Hamiltonian families \cite{Carrasquilla_2020, carleo2017solving, Carleo_2019, lewis2023improvedmachinelearningalgorithm, huang2022provably}. This development is driven by the fundamental necessity of many-body analysis in condensed matter physics, quantum chemistry, and materials science, where researchers frequently require repeated solutions for ensembles of closely related Hamiltonians. Formally, the goal is to learn ground-state properties over a parametric family $H(x)$, where $x$ represents real-valued parameters such as nuclear configurations in potential energy surfaces, coupling strengths in lattice models, or external fields in phase diagrams. A critical opportunity in this domain lies in the structure of the Hilbert space. Although the total dimension scales exponentially with the number of particles $n$, physically realizable ground states typically inhabit a low-dimensional manifold constrained by locality, symmetry, smooth parameter dependence, and restricted entanglement patterns. This suggests that the mapping from Hamiltonian parameters to physical observables admits efficient, learnable representations that can bypass the curse of dimensionality. Furthermore, while conventional numerical methods, such as exact diagonalization (ED) \cite{weisse2008exact}, density matrix renormalization group (DMRG) \cite{white1992density}, and quantum Monte Carlo (QMC) \cite{nightingale1998quantum}, as well as quantum-coherent algorithms like quantum phase estimation (QPE) \cite{cleve1998quantum}, provide high precision for a single instance $H(x)$, they are strictly instance-specific. ML-based approaches shift this paradigm toward amortized computation. By leveraging a representative training set of previously computed ground states, these models learn to interpolate and generalize across the parameter manifold. This strategy effectively \textit{front-loads} the computational complexity into a one-time training phase. Once optimized, the model enables simpler inference for unseen parameter points, dramatically reducing the cumulative overhead required to map out physical landscapes. Consequently, machine learning transforms repeated, expensive simulations into a predictive inference problem, exploiting shared geometric and statistical structures to bypass the limitations of traditional many-body solvers.}

{In the quantum setting, variational quantum algorithms (VQAs) have emerged as the primary candidates for implementing such learnable solvers on near-term devices. These algorithms typically operate in a hybrid quantum-classical framework, where quantum computers evaluate the cost function, and classical optimizers train a parameterized quantum circuit to minimize this cost~\cite{cerezo2021variational, peruzzo2014variational}. Their flexibility and compatibility with noisy intermediate-scale quantum devices make them particularly attractive for approximating ground states of quantum many-body problems. However, although there are numerous empirical studies~\cite{kandala2017hardware, belaloui2024groundstateenergyestimation, Lolur_2021, de_Keijzer_2022}, rigorous performance guarantees for VQAs remain scarce~\cite{schatzki2024theoretical}. Current theoretical literature has focused heavily on \textit{trainability}. It is well-documented that generic quantum neural networks often encounter barren plateaus, where gradients vanish exponentially with problem size \cite{mcclean2018barren, cerezo2021cost, marrero2021entanglementinducedbarrenplateaus}. This phenomenon casts significant doubt on the scalability of VQAs to classically intractable domains \cite{mcclean2018barren, Ragone_2024}. Even in the absence of barren plateaus, the optimization landscape is frequently swamped by a proliferation of poor local minima \cite{anschuetz2022beyond}. While these challenges are formidable, the studies also suggest they may be circumvented through good initialization \cite{anschuetz2022beyond, mcclean2018barren}, giving rise to the theory of warm-starting in VQAs \cite{mhiri2025unifyingaccountwarmstart}.}

\begin{figure}[!hbt]
    \centering
     \includegraphics[width=0.9\textwidth]{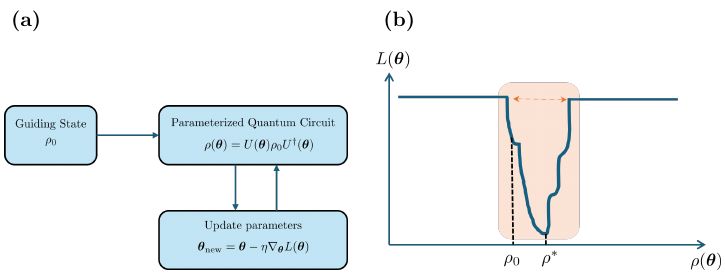}
    \caption{}
    \label{fig: problem_fig}
\end{figure}

{However, existing research on warm-starting \cite{mhiri2025unifyingaccountwarmstart,drudis2024variational, meyer2024warmstartvariationalquantumpolicy, egger2021warm} has primarily focused on parameter initialization. These strategies involve sampling parameters from specific probability distributions \cite{zhang2022escaping, park2024hamiltonian, park2024hardware, shi2024avoidingbarrenplateausgaussian} or transferring optimized parameters from smaller problem instances \cite{brandao2018fixed, farhi2022quantum}. While effective at smoothing the optimization landscape, these methods leave a critical theoretical gap: they do not address whether warm-started VQAs generalize effectively. In this work, we shift the focus from trainability to learnability, aiming to provide a comprehensive analysis of both convergence and generalization. Rather than treating warm-starting as a parameter choice, we define it through the lens of state initialization, assuming an initial state that maintains a non-trivial overlap with the target ground state (see Figure~\ref{fig: problem_fig}a).} 

{This perspective is inspired by standard quantum phase estimation (QPE) methods, which require a \textit{guiding state} with non-negligible overlap to achieve high-precision energy estimation. This reveals a fundamental trade-off: QPE provides rigorous guarantees but remains impractical for the NISQ era \cite{whitfield2013computational, dalzell2023quantum}, whereas VQAs offer near-term practicality but lack formal performance bounds. Since both paradigms ultimately rely on the availability of a guiding state, we address the following central question:
\begin{center}
\textit{Can we have guarantees in VQAs with the guiding state assumption?}
\end{center}
Our paper addresses this question by designing a VQA architecture that enables us to establish convergence and generalization results under the guiding state assumption. To prove our results, we develop a \textit{linearization trick} that may be of independent interest. In particular, the trick allows for the analysis of convergence and generalization of geometrically local circuits under a good initialization via a formal correspondence between the non-linear training dynamics of the variational circuit and the evolution of a fixed-kernel model, akin to the \textit{neural tangent kernel} (NTK) in classical deep learning~\cite{jacot2020neuraltangentkernelconvergence}.}

{In particular, we consider the problem of estimating ground-state properties of a parametric class of \textit{local Hamiltonians}, $H(x)$, which depends smoothly on the real-valued parameters $x \in \mathcal{X}$. with respect to a known $k$-local observable $O$. Concretely, we consider a supervised learning setting in which the model is trained on a dataset $\mathcal{S} = \{(x_i, y_i)\}_{i=1}^M$, where $y_i \approx \Tr[O \rho(x_i)]$ denotes the ground-state property corresponding to $x_i$, with samples drawn from a distribution over $\mathcal{X}$. The goal is then to learn a model that, once trained, can accurately predict ground-state properties for previously unseen inputs $x'$ under the observable $O$, thereby amortizing the computational cost across the parameter space. In practice, the training dataset $\mathcal{S}$ could be obtained from either classical simulations or quantum experiments. These labels may be obtained either from trusted classical simulations on a restricted subset of tractable instances, or from prior quantum experiments in which the observable $O$ is measured on systems prepared at selected parameter values. This setting is natural in many-body physics and quantum chemistry, where high-quality information is often available only for a sparse collection of systems, and exhaustively simulating or measuring every new instance is prohibitively costly. Our goal is therefore not to assume free access to ground-state labels for all inputs, but rather to study the practically relevant amortized regime in which information gathered from previously solved or measured instances is reused to predict properties of previously unexplored Hamiltonians, avoiding the need for a completely new expensive simulation or deeper quantum subroutines whenever a new parameter setting is encountered.}

{Inspired by the guiding-state paradigm underlying QPE, we assume access to an oracle that, given $x$, outputs a guiding state $\rho_0(x)$ that is $\delta$-close to the ground state $\rho(x)$, as shown in Figure~\ref{fig: problem_fig}b. This assumption is introduced not as a claim about all physically relevant settings, since finding and preparing guiding states for generic Hamiltonians is itself widely believed to be computationally hard~\cite{waite2025physicallymotivatedguidingstateslocal}. Rather, it is intended to formalize a restricted but practically meaningful warm-start regime in which prior heuristic, numerical, or experimental information provides a nontrivial initialization for the learning problem. Further discussion on some methods of preparing the guiding state is presented in the Methods section. }

{ In this work, we introduce a VQA architecture using \textit{alternating layered ansatz} (ALA)~\cite{nakaji2021expressibility} with guiding state inputs and Gaussian parameter initialization near zero. The central contribution of this work is providing a theoretical tool: \textit{linearization trick} to analyze convergence and generalization of geometrically local circuits under a guiding state assumption. The trick largely relies on a concentration phenomenon: in the infinite-system limit, the evolution of the variational circuit aligns with that of a fixed kernel, obtained by linearizing the model around its initialization. This establishes a direct correspondence between guiding state VQAs and kernel methods, akin to the neural tangent kernel (NTK) in classical deep learning \cite{jacot2020neuraltangentkernelconvergence}. Leveraging tools from kernel theory \cite{shawe2004kernel} and statistical learning theory \cite{shalev2014understanding}, we derive the first guarantees on both convergence and generalization error for VQAs under the guiding state assumption. }

{First, we establish a rigorous convergence guarantee (Theorem~\ref{thrm: convergence_paper_ver}) showing that the training loss decays at a rate fundamentally governed by the spectrum of the quantum neural tangent kernel (QNTK). In particular, the error contracts most rapidly along the dominant eigenspaces of the kernel, while the residual correction term is explicitly suppressed by the quality of the guiding state initialization. Second, we derive a generalization bound for the VQAs (Theorem~\ref{thrm: main_theorem}), proving that the generalization error scales as $\tilde{\mathcal O}(B\delta/\sqrt{M})$, where $B$ depends on the QNTK spectrum. These bounds reveal that predictive performance is jointly controlled by two central quantities: the spectral structure of the induced kernel and the overlap quality of the guiding state. To verify our theoretical framework, we perform experiments on two-dimensional anti-ferromagnetic Heisenberg models. Our results show that even with 20 qubits, the dynamics of the guiding-state VQA closely align with its linearized kernel counterpart, while the observed generalization error scales consistently with the growth of the training dataset.}

\section*{Results}

\subsection*{Variational Quantum Algorithm with Guiding States} \label{sec: task_spec}


We consider the problem of predicting ground state properties of a family of local Hamiltonians $H(x) \in \mathbb{C}^{2^n\times 2^n}$ that is configured by $m$ real parameters $x\in \mathcal{X}$. It can be the coupling constant vectors of Ising models on a fixed lattice or classical nuclear coordinates in electronic molecular Hamiltonians. The goal is to learn a model to predict the properties of ground state $\rho(x)$ from a known observable $O$. In analogy with the QPE setting, we assume access to an oracle that, given an input parameter $x$, outputs a \emph{guiding state} $\rho_0(x)$. This guiding state serves as an approximation to the true ground state and satisfies
\begin{equation}
d_{\Tr}(\rho_0(x), \rho(x)):= \frac{1}{2}\Vert \rho_0(x) - \rho(x) \Vert_{1} \leq \delta \quad \forall x ,
\label{eqn: guided_state}
\end{equation}
where $\delta$ quantifies the gap in trace distance. In the context of guided local Hamiltonians, it is common to assume $\delta \in \mathcal{O}(1/\text{poly}(n))$~\cite{gharibian2022improved, gharibian2022dequantizing}. This assumption is especially relevant to quantum chemistry problems whereby the observation that, in practice, computationally efficient classical techniques often give a fairly good initial approximation of the ground state. In this work, we will keep this term general. It could be either a constant or decay with system size $n$. In addition, although they commonly use fidelity as the measure of closeness between $\rho_0(x)$ and $\rho(x)$, this could be easily bounded by the trace distance, ensuring that the established results for guided local Hamiltonian remain valid in our setting.
\begin{figure}[!hbt]
    \centering
    \includegraphics[width=0.9\textwidth]{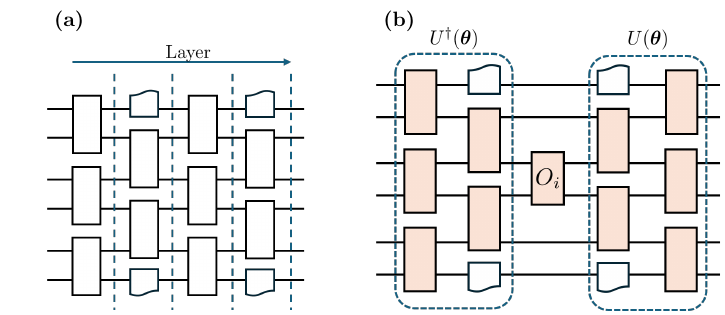}
    \caption{} 
    \label{fig: ala}
\end{figure}
We consider a supervised learning setting in which the algorithm is provided with a finite collection of problem instances drawn from a distribution $\mathcal{D}$ over a parameter space $\mathcal{X}$. Each instance $x \in \mathcal{X}$ specifies a Hamiltonian $H(x)$ with ground state $\rho(x)$, together with a target quantity given by the expectation value of a fixed observable $O$. Concretely, the training data take the form
\begin{equation}
\mathcal{S} = \{(x_i, y_i)\}_{i=1}^{M},
\label{eqn: training_dataset}
\end{equation}
where $x_i \sim \mathcal{D}$ and $y_i = \Tr[O \rho(x_i)]$. Such data may arise either from classical simulations (when available) or from prior quantum experiments. After training, the algorithm is used to predict $\Tr[O\rho(x')]$ for previously unseen inputs $x' \sim \mathcal{D}$.

The model is defined by a parameterized quantum circuit $U(\boldsymbol{\theta})$ acting on an input-dependent guiding state $\rho_0(x)$. The resulting hypothesis is
\begin{equation}\label{eqn: model_function}
f_{\boldsymbol{\theta}}(x) := \Tr\!\left[ O \, U(\boldsymbol{\theta}) \rho_0(x) U^{\dagger}(\boldsymbol{\theta}) \right],
\end{equation}
and the parameters $\boldsymbol{\theta}$ are optimized by minimizing the empirical squared loss~\eqref{eqn: loss_function}. We focus on observables that admit a local decomposition,
\begin{equation}
O = \frac{1}{K}\sum_{\ell=1}^{K} O_\ell ,
\label{eqn: observables}
\end{equation}
where each $O_\ell$ acts non-trivially on at most $k$ qubits and satisfies $\|O_\ell\|_{\mathrm{op}} \leq 1$. This setting encompasses most physically relevant observables in quantum many-body systems, where interactions and measurements are typically local. The normalization ensures $\|O\|_{\mathrm{op}} \leq 1$ and simplifies subsequent bounds.

Finally, our analysis relies on a mild smoothness condition on the parameterized circuit, namely that the state $U(\boldsymbol{\theta})\rho_0(x)U^{\dagger}(\boldsymbol{\theta})$ varies smoothly with respect to the parameters. Specifically, we require the existence of a constant $c>0$ such that
\begin{equation}
\left\Vert \frac{\partial}{\partial \theta_j} 
U(\boldsymbol{\theta}) \rho_0(x) U^{\dagger}(\boldsymbol{\theta}) \right\Vert_2,
\quad
\left\Vert \frac{\partial^2}{\partial \theta_i \partial \theta_j} 
U(\boldsymbol{\theta}) \rho_0(x) U^{\dagger}(\boldsymbol{\theta}) \right\Vert_2
\;\leq\; c ,
\label{eqn: assumption}
\end{equation}
uniformly over all parameters and inputs. This condition is standard in analyses of variational circuits and is naturally satisfied by shallow, locality-preserving ansätze, including the alternating layered ansatz considered below.

The guarantee of a variational quantum algorithm depends on a variety of factors, including the choice of ansatz $U(\boldsymbol{\theta})$. In this work, we particularly employ alternating layered ansatz (ALA), which has both important properties of trainability and expressibility \cite{nakaji2021expressibility}. The ALA introduced in \cite{cerezo2021cost} consists of multiple layers, each of them having some separated blocks that have parameterized single-qubit rotations and a fixed entanglement layer connecting all qubits inside the block. An illustration of the ALA is shown in Figure~\ref{fig: ala}a. For simplicity, we further assume that each block contains an even number, $m$, qubits ($m$ is independent of $n$), and $n/m$ is an integer. That means in the odd-numbered layer, there are $n/m$ blocks which non-trivially act on $\{1,..., m\}, \{m+1,..., 2m\}, ..., \{n-m+1,..., n\}$ qubits, while the even-numbered layer contains $n/m+1$ blocks which acts on $\{1,..., m/2\}, \{m/2+1,..., 3m/2\}, ..., \{n-m/2+1,..., n\}$ (the first and the last blocks operate on $m/2$ qubits). In detail, we define the ALA used in our result in Definition~\ref{def: ALA}.  

\begin{defn}\label{def: ALA}
    An $n$-qubit unitary $U(\boldsymbol{\theta})$ is ALA($n, m, p, L$) if it consists of $L$ layer where the odd-numbered layer is expressed as:
    \begin{equation}
        U_{\text{odd}}(\boldsymbol{\theta}) = \prod_{i=0}^{n/m-1} W_{[mi+1, m(i+1)]}(\boldsymbol{\theta})_i
    \end{equation}
    and the even-numbered layer is defined as:
    \begin{equation}
         U_{\text{even}}(\boldsymbol{\theta}) = W_{[n-m/2+1, n]}(\boldsymbol{\theta})_{n/m+1}\prod_{i=0}^{n/m-2} W_{[m(i+1/2)+1, m(i+3/2)]}(\boldsymbol{\theta})_i.W_{[1, m/2]}(\boldsymbol{\theta})_1
    \end{equation}
    where $W_{[a,b]}(\boldsymbol{\theta})$ contains single-qubit rotations and entangler acting on from $a^{th}$ to $b^{th}$ qubits parameterized by $p$ parameters.
\end{defn}
The advantage of the ALA lies in its co-existence of trainability and expressibility. According to Cerezo et al. \cite{cerezo2021cost}, an instance of ALA($n,m,p,L$) avoids the issue of barren plateaus if it meets the following criteria: (i) the cost function is defined as a sum of few-body operators, and (ii) the number of layers $L$ scales logarithmically with the system size, specifically $L\in \mathcal{O}(\log(n))$. Remarkably, recent findings indicate that shallow ALA can achieve high expressibility when the ensemble of unitary matrices in each block forms a 2-design \cite{nakaji2021expressibility}. To fulfill this requirement, it is necessary for each block to contain $p\in \mathcal{O}(m)$ parameters to approximate a one-dimensional 2-design \cite{brandao2016local, harrow2023approximate}, while only $\mathcal{O}(\sqrt{m})$ parameters are needed for two-dimensional connectivity \cite{harrow2023approximate}. Notably, these parameter requirements do not depend on the total number of qubits $n$. Therefore, the ALA enables us to leverage both trainability and expressibility effectively, even when employing a shallow depth circuit. 

Therefore, it is natural to consider 
\( U(\boldsymbol{\theta}) \in \text{ALA}(n,m,p,L) \) with \( m, L, p = \mathcal{O}(\log(n)) \). Under this setting, the model functions generated by ALA exhibit a crucial locality property: the measurement of an observable \( O_i \) depends only on the light cone of the qubits on which \( O_i \) acts, as illustrated in Figure~\ref{fig: ala}b. 
A formal proof of this statement will be given later (Lemma~\ref{lmma: locality}). 
This locality property plays a central role in our main result, 
which will be established in a subsequent section.

{More broadly, our proof technique generally applies to any variational ansatz satisfying this same locality condition. 
Thus, rather than focusing on problem-specific ansatz design, we keep the framework general and do not impose any explicit dependence on the Hamiltonian family \( H(x) \). 
On the one hand, ALA is sufficiently expressive to define a broad variational hypothesis class while still retaining the trainability and locality properties required for our analysis. 
On the other hand, this choice does not preclude the use of more problem-informed circuit constructions. 
For instance, one may consider Hamiltonian variational ans\"atze (HVA)~\cite{park2024hamiltonian} tailored to the structure of \( H(x) \). 
For many commonly studied local Hamiltonians, the resulting layered time-evolution circuits induce the same bounded light-cone structure captured by ALA. 
From this perspective, ALA should be viewed as a sufficiently general framework that encompasses many local Hamiltonian-inspired constructions while remaining architecture-agnostic enough to support a family-level learning analysis.}


For parameter initialization, we leverage the guiding states, which give us access to a `good' approximation of the ground state. Instead of randomly initializing the parameters, which could cause the output state to move arbitrarily across the Hilbert space, we initialize the parameters near the vector $\boldsymbol{0}$. Specifically, we model the initialized parameters as drawn from a Gaussian distribution:
\begin{equation}
    \boldsymbol{\theta}(0) \sim \mathcal{N}(\boldsymbol{0}, \kappa^2\mathbf{I}) 
    \label{eqn: param_init}
\end{equation}
where $0 < \kappa \leq 1$ controls the magnitude of initialization, and all randomnesses are independent. This initialization strategy serves two key purposes. First, this avoids the output state being distributed randomly due to the high expressibility of ALA. Second, it allows us to take advantage of the guide state that is already close to the ground state, so this initialized scheme will put the output state into a more favorable region near the optimal value.

Overall, our algorithm is then described as follows:
\begin{algorithm}[H]
\caption{Variational Quantum Algorithm with guiding states}
\begin{algorithmic}[1]
\State \textbf{Input:} 
\begin{itemize}
    \item Training data $\mathcal{S}$~\eqref{eqn: training_dataset} defined on a system size $n$
    \item Observable $O$~\eqref{eqn: observables}
    \item Circuit configurations $m, p, L$, which are $\mathcal{O}(\log(n))$
    \item Learning rate $\eta$
    \item Number of iterations $T$
\end{itemize}
\State \textbf{Output:} Trained model $\boldsymbol{\theta}$

\State Initialize the ALA($n, m, p, L$) circuit with parameters $\boldsymbol{\theta}(0) \sim \mathcal{N}(\boldsymbol{0}, \kappa^2 \mathbf{I})$
\For{t = 1 to $T$}
    \For{i = 1 to $M$} \Comment{M is the number of training samples in $\mathcal{S}$}
        \State $\hat{y}_i \gets \text{Forward the circuit with } \rho_0(x_i)$
    \EndFor
    \State $L_{\mathcal{S}}(\boldsymbol{\theta}) \gets \text{Compute loss function as~\eqref{eqn: loss_function}}$
    \State $ \nabla_{\boldsymbol{\theta}} L_{\mathcal{S}}(\boldsymbol{\theta}) \gets \text{Compute gradients of } \boldsymbol{\theta} \text{ w.r.t $L_{\mathcal{S}}(\boldsymbol{\theta})$}$
    \State $\boldsymbol{\theta}(t+1) \gets \boldsymbol{\theta}(t) - \eta \nabla_{\boldsymbol{\theta}}L_{\mathcal{S}}(\boldsymbol{\theta})$ \Comment{Update parameters}
\EndFor
\State \Return $\boldsymbol{\theta}$
\end{algorithmic}
\label{alg: main_alg}
\end{algorithm}
\subsection*{Theoretical Guarantee}
Next, we present the main results of our study, focusing on both the convergence properties and generalization performance of Algorithm~\ref{alg: main_alg}. We first analyze the convergence behavior, followed by a detailed study of its generalization capability.

The central tool in our analysis is a \textit{linearization trick}, which maps the nonlinear training dynamics induced by the model in~\eqref{eqn: model_function} to an approximately linear dynamical system. In particular, we leverage the neural tangent kernel (NTK) framework for classical neural networks, where the training dynamics are governed by the $M \times M$ kernel matrix
\begin{equation}
    K_{\boldsymbol{\theta}}(x,x') := \frac{1}{M}\left\langle \nabla_{\boldsymbol{\theta}} f_{\boldsymbol{\theta}}(x), \nabla_{\boldsymbol{\theta}} f_{\boldsymbol{\theta}}(x') \right\rangle.
\end{equation}
In standard overparameterized regimes, this kernel concentrates around a deterministic limit, yielding stable and well-conditioned optimization dynamics in classical neural networks. However, such concentration behavior does not generally extend to quantum circuits~\cite{liu2023analytic}. Motivated by this limitation, and by recent work characterizing conditions under which quantum neural networks enter a lazy training regime~\cite{abedi2023quantum}, we depart from the classical parameter-to-data scaling perspective. Instead, we establish kernel concentration through the \textit{locality structure} of the quantum ansatz. This allows us to reestablish NTK concentration for variational quantum algorithms under geometrically local architectures, corresponding to the setting considered in Algorithm~\ref{alg: main_alg}. As a consequence, the training dynamics of the model in~\eqref{eqn: model_function} under gradient descent are well-approximated by those of a linear model given by the first-order Taylor expansion around initialization. The formal definitions, along with detailed proofs of NTK concentration in Algorithm~\ref{alg: main_alg}, are provided in the Methods section.

Building on this characterization, our analysis combines standard tools from statistical learning theory~\cite{shalev2014understanding} and kernel methods~\cite{shawe2004kernel} to study both the convergence and generalization properties of the resulting linearized model. We then transfer these guarantees back to the original nonlinear model, with additional correction terms that quantify the approximation error introduced by the linearization. In the subsequent results, we further show that these correction terms can be controlled by the quality of the guiding states.

The key insight in our convergence analysis is that, conditioned on the guiding states, only a small initialization scale $\kappa$ is required to establish concentration of the quantum neural tangent kernel for Algorithm~\ref{alg: main_alg}. Here, we denote by $K_{\boldsymbol{\theta}(0)}$ the NTK evaluated at random initialization. Let $\{\lambda_j\}_j$ denote its eigenvalues, and define
$
\lambda_{\min} := \lambda_{\min}(K_{\boldsymbol{\theta}(0)})$, $
\lambda_{\max} := \lambda_{\max}(K_{\boldsymbol{\theta}(0)})$.
The resulting convergence analysis is as follows:
\begin{thrm}[Convergence]
    Under the training procedure in Algorithm~\ref{alg: main_alg}, suppose {$0 < \lambda_{\min}:= \lambda_{\min}(K_{\boldsymbol{\theta}(0)}) \leq \lambda_j \leq \lambda_{\max} := \lambda_{\max}(K_{\boldsymbol{\theta}(0)}) < \infty$} and for $\eta \leq \frac{1}{\lambda_{\max}}$, $\kappa = \Tilde{\mathcal{O}}(\frac{\delta\sqrt{\gamma}}{n})$. Then, with probability at least $1 - \gamma$ over the random initialization, {we have for all \(t> 0\)}:
    \begin{equation}
        L_{\mathcal{S}}(\boldsymbol{\theta}(t)) = \Tilde{\mathcal{O}}\left(\sum_{j} (1-\eta \lambda_{j})^{2t}\delta^2+\frac{n}{K^3}\eta^2 t^2 \delta^3\right)
    \end{equation}
    \label{thrm: convergence_paper_ver}
\end{thrm}
In light of this result, the dominant term $\sum_{j} (1-\eta \lambda_{j})^{2t}\delta^2$ vanishes as $t$ increases. Convergence is faster along directions associated with larger eigenvalues $\lambda_j$, consistent with findings in the classical neural tangent kernel (NTK) setting \cite{jacot2020neuraltangentkernelconvergence}. Intuitively, larger eigenvalues correspond to directions in parameter space where the loss surface has steeper curvature, leading to more rapid error reduction. This behavior mirrors the dynamics of linear models and the NTK approximation, where convergence is primarily governed by the learning rate $\eta$ and the spectrum of the tangent kernel. In contrast, directions associated with smaller eigenvalues converge more slowly, producing a staggered convergence pattern across different parameter dimensions.  

The correction term $\tfrac{n}{K^3}\eta^2 t^2 \delta^3$, by contrast, grows quadratically with $t$. However, this growth is substantially mitigated by the denominator $K^3$, which scales polynomially with $n$, ensuring that the contribution of this term is increasingly suppressed as the system size grows. Moreover, the use of guiding states reduces the number of update steps required, further controlling its magnitude. {Importantly, $t$ is never allowed to grow arbitrarily large: as established later in Theorem~\ref{thrm: main_theorem}, the total training horizon satisfies
{$
T=\mathcal{O}\!\left( \frac{\ln M}{\eta\lambda_{\min}} \right).
$}
Therefore, as long as $\lambda_{\min}$ does not decay too rapidly with the problem size (e.g., remains at least inverse-polynomial), this correction term remains bounded and does not compromise the convergence guarantee. Consequently, the overall convergence behavior is primarily determined by the spectral properties, ensuring both stability and scalability of the variational quantum algorithm. These results also offer guidance for investigating the inductive bias of ansätze through the spectral properties of their associated quantum tangent kernel, an intriguing open question for future research.}

We next provide the generalizability of the learning model in Algorithm~\ref{alg: main_alg}. In particular, we study the theoretical bound for the \textit{generalization error}, which is defined as:
\begin{equation}
    \text{gen}(\boldsymbol{\theta}) := \left| \mathbb{E}_{x\sim \mathcal{D}}[ \left| f_{\boldsymbol{\theta}}(x) - \Tr[O\rho(x)]\right|] - \mathbb{E}_{x\sim \mathcal{S}}[ \left| f_{\boldsymbol{\theta}}(x) - \Tr[O\rho(x)]\right|]\right|. 
    \label{eqn: gen_def}
\end{equation}
The following theorem establishes the theoretical bound of the generalization error of the Algorithm~\ref{alg: main_alg}:     
\begin{thrm}[Generalization]
    Consider a training dataset $\mathcal{S}$~\eqref{eqn: training_dataset} defined on a system size $n$ with the observable $O$ is represented as~\eqref{eqn: observables}. If the Algorithm~\ref{alg: main_alg} is trained on $\mathcal{S}$ with $\eta \leq \frac{1}{\lambda_{\max}}$ and $\kappa = \Tilde{\mathcal{O}}(\frac{\delta\sqrt{\gamma}}{n})$. Suppose {$0 < \lambda_{\min}:= \lambda_{\min}(K_{\boldsymbol{\theta}(0)}) \leq \lambda_j \leq \lambda_{\max} := \lambda_{\max}(K_{\boldsymbol{\theta}(0)}) < \infty$}. {For fixed constants $0<B_1< B_2<\infty$ and let $T$ be any finite training iteration satisfying $B_1\leq\Tr\!\left[(I-\eta K_{\boldsymbol{\theta}(0)})^T\right]\leq B_2$. Any such iteration satisfies $T= \mathcal O\left(\frac{\ln M}{\eta\lambda_{\min}}\right)$, following inequality holds with probability at least $1-\gamma$
    \begin{equation}
        \text{gen}(\boldsymbol{\theta}(T)) \leq \Tilde{\mathcal{O}}\left(B_2\delta\sqrt{\frac{1}{M}} + \sqrt{\frac{\ln(1/\gamma)}{M}} + {\frac{n}{K^3} \left(\frac{\ln(M/B_1)}{\lambda_{\min}}\right)^2 \delta^2} \right).
    \end{equation}
    }
    \label{thrm: main_theorem}
\end{thrm}
Now, we discuss our generalization bound. The dominating term is:
\begin{equation}
    B_2\delta\sqrt{\frac{1}{M}}.
\end{equation}
The value of $B_2$ plays a crucial role in controlling the generalization error and indicates the behavior of the trace of the discrete transition matrix, $(\mathbf{I} - \eta K_{\boldsymbol{\theta}(0)})^T$. $B_2$ indeed provides an upper bound on the trace of the matrix that governs the convergence behavior of the algorithm. Consequently, focusing on the leading statistical terms in the generalization error $\text{gen}(\boldsymbol{\theta}(T)) \leq \epsilon$ requires a sample complexity of $\mathcal{O}\left(\frac{B^2_2\delta^2 + \log(1/\gamma)}{\epsilon^2}\right)$.

On the other hand, the model’s efficiency in general settings is strongly influenced 
by the eigensystem of the kernel \(K_{\boldsymbol{\theta}(0)}\). 
In the worst case, this may require an impractically large amount of training data for good performance. 
However, this requirement can be substantially reduced by selecting an appropriate kernel. 
The central challenge lies in identifying a kernel that effectively encodes the relevant structure of the problem. 
Kernel design reflects how prior knowledge is embedded and exploited, 
yet determining the right inductive bias is often nontrivial \cite{shalev2014understanding}. 
The difficulty stems from the need to express the inductive bias that best aligns with the data’s underlying structure. 
Several works have investigated ways to incorporate such inductive bias into quantum kernels 
\cite{schatzki2024theoretical, kubler2021inductive, liu2021rigorous}. 
We further remark that Theorem~\ref{thrm: main_theorem} suggests an alternative route to inducing inductive bias---through the spectrum of the tangent kernel. 
If this spectral bias can be properly controlled, it has the potential to accelerate both training and generalization.

We again emphasize the role of guiding states in our algorithm. The quality of the guiding state, captured by the parameter $\delta$, directly influences the generalization bound through the additional term  
\begin{equation}
    {\frac{n}{K^3}\left(\frac{\ln(M/B_1)}{\lambda_{\min}}\right)^2 \delta^2}.
\end{equation}
This contribution is most relevant in regimes where the system size $n$ is finite but large. While the factor $\frac{n}{K^3}$ ensures that its impact diminishes as the system grows, for smaller $n$ the term can dominate if the guiding state is of poor quality (i.e., $\delta$ is large). The quadratic dependence on $\delta$ underscores how strongly the accuracy of the guiding state affects the generalization behavior. In the context of guided local Hamiltonian problems, where $\delta = \mathcal{O}(1/\mathrm{poly}(n))$, this effect becomes particularly favorable. High-quality guiding states suppress the additional term, making the bound more favorable even at modest system sizes.

\begin{figure}[!hbt]
    \centering
    \includegraphics[width=\textwidth]{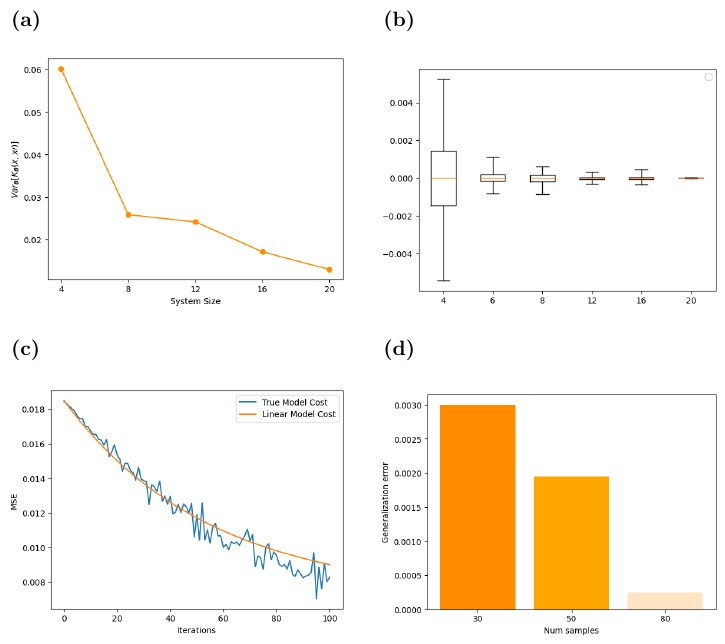}
    \caption{}
    \label{fig:num_exp}
\end{figure}

\begin{figure}
    \centering
    \includegraphics[width=0.9\textwidth]{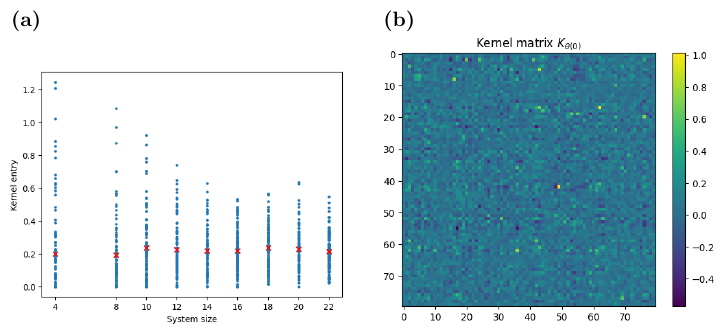}
    \caption{}
    \label{fig:num_exp2}
\end{figure}
\subsection*{Numerical Results} \label{sec: experiments}

Here, we present numerical experiments to assess the performance of our algorithm in practice. Here, we consider the two-dimensional anti-ferromagnetic Heisenberg model with Hamiltonian of:
\begin{equation}
    H(x) = \sum_{\langle ij \rangle} x_{ij}(X_iX_j + Y_iY_j + Z_iZ_j)
\end{equation}
where the sum is over the nearest neighbors in a 2D lattice, which accounts for the interaction between each pair of adjacent spins. The values of $x = (x_{ij})_{i,j}^n$ represent the coupling matrix determining the strength of the interactions. For any observable $O$, the goal is using Algorithm~\ref{alg: main_alg} to produce a hypothesis $f_{\boldsymbol{\theta}}$ such that:
\begin{equation}
    f_{\boldsymbol{\theta}}(x) = \bra{\psi(x)}O\ket{\psi(x)}
\end{equation}
where $\ket{\psi(x)}$ denotes the ground state of $H(x)$. In these experiments, we focus on the Hamiltonian $H(x)$ on a $2\times 10$ lattices. The dataset is generated by uniformly sampling the coupling constant $\{x_{ij}\}$ at random from the interval $[0,2]$. We train our Algorithm~\ref{alg: main_alg} on a training dataset of $80$ randomly chosen values of $x=\{x_{ij}\}$ and validate on a $20$-sample testing dataset. The algorithm will aim to predict the ground state properties of $O = \frac{1}{{n}}\sum_{i=1}^{n} Z_i$, where $n$ is the number of qubits. In each sample, the guiding state is generated as follows:
\begin{equation}
    \ket{\psi_0(x)} = \sqrt{1-\delta^2}\ket{\psi(x)} + \delta \ket{\psi^{\perp}(x)}
\end{equation}
where $\ket{\psi^{\perp}(x)}$ is one of the excited states of the Hamiltonian $H(x)$ and $\delta$ defines the closeness between the initial state and the ground state, which is set as $1/n^2$. 

In the following, we characterize our theoretical claims. These first include analyzing the convergence of the quantum neural tangent kernel in our construction and learning behaviors of the true model with respect to the linearized model governed by the initialized tangent kernel. Then, we validate the generalization error of our algorithm. 

In Figure~\ref{fig:num_exp}a, we show the variance of a single kernel entry at over $100$ random initialization of $\boldsymbol{\theta}(0)$. We can see that as the system size increases, the deviation of the tangent kernel entry decreases, indicating that the kernel entries become more concentrated around their mean. To demonstrate the convergence of the quantum neural tangent kernel, we further illustrate the amount of {$\Delta_t K_{\boldsymbol\theta}(x,x') := K_{\boldsymbol \theta(t+1)}(x,x')-K_{\boldsymbol\theta(t)}(x,x')$} over the range of $t$ from 1 to 100 when training with the Algorithm~\ref{alg: main_alg} in Figure~\ref{fig:num_exp}b. The figure shows the distribution of the updates of elements in the tangent kernel at different system sizes. We can see that the median (orange line) is close to zero for all system sizes, indicating that the central tendency of the data is near zero, regardless of the system size. As the system size increases, the width of the boxes tends to decrease, meaning the values become more concentrated around the mean, which is close to zero. This suggests that the values of {$\Delta_t K_{\boldsymbol\theta}(x,x')$}tend to approach zero as the system size approaches infinity.

The first two experiments help us to see clearly the convergence of the quantum neural tangent kernel that will support us in analyzing the performance of our true model through the linearized model. In particular, we are interested in the training and generalization error of our algorithm in Figures~\ref{fig:num_exp}c and~\ref{fig:num_exp}d. For these experiments, we work with the system size of $20$ and train the algorithm with the alternating layered ansatz of $m=4$ and $L=2$ and run with 100 training iterations. First, we analyze the asymptotical training behavior of the model compared to the linearized model. In Figure~\ref{fig:num_exp}c, we can see that the training error of the model asymptotically behaves similarly to its linearized version, demonstrating our linearization trick. That means when the system size is large enough, the performance of the linearized model could well-characterize the true model. Next, we study the generalization error of the model training with our Algorithm~\ref{alg: main_alg} for different training dataset sizes $M$. As seen in Figure~\ref{fig:num_exp}d, the generalization error drops significantly with the number of training samples $M$.

Theoretical analysis shows that different eigenspaces of the initialized tangent kernel \(K_{\boldsymbol{\theta}(0)}\) are learned at different rates. As established in Theorem~\ref{thrm: convergence_paper_ver}, the contribution associated with an eigenvalue \(\lambda_j\) decays according to
\((1-\eta\lambda_j)^{2t}\). Hence, directions corresponding to larger eigenvalues are fitted more rapidly, whereas components associated with smaller eigenvalues require a longer training horizon. In the numerical experiments of Figure~\ref{fig:num_exp}d, we use a fixed finite training horizon \(T=100\)
with learning rate \(\eta=10^{-3}\). This induces an early-stopping-like effect: within a finite number of iterations, the dynamics preferentially resolve the dominant eigendirections of the tangent kernel, while directions associated with smaller eigenvalues remain comparatively less fitted.

To illustrate the spectral structure underlying this behavior, Figure~\ref{fig:num_exp2} displays the empirical tangent kernel and its
spectrum. We emphasize that the kernel plotted in this figure is the \emph{unscaled} empirical NTK, computed without the \(1/M\) normalization and in the theoretical analysis. Consequently, the trace and eigenvalues reported in Figure~\ref{fig:num_exp2} are larger by a factor of \(M\) than those of the normalized theoretical kernel. In particular, at \(M=80\), the unscaled kernel has $\operatorname{Tr}[K_{\boldsymbol{\theta}(0)}]\approx 16$ and a dominant eigenvalue of approximately \(5.5\), as shown in Figure~\ref{fig:num_exp2}b. Along this dominant eigenspace, the reduction factor after \(100\) discrete gradient-descent steps is \( (1-\eta\lambda)^{2T} = (1-10^{-3}\times 5.5)^{200} \approx 0.33\). This signifies at least a $67\%$ reduction in error for the most relevant component.

\section*{Discussion}
This work introduces a guiding-state variational quantum algorithm for predicting ground-state properties and develops a proof framework based on linearization to characterize its learning dynamics. By establishing a concentration property of the training trajectory, we show that the nonlinear variational optimization can be accurately approximated by a kernel model, leading to rigorous guarantees on both convergence and generalization. A central insight of our analysis is the role of guiding states in shaping the finite-size learning behavior of the model. In particular, high-quality guiding states accelerate convergence, suppress finite-size correction terms, and improve optimization stability across system dimensions. Although their explicit contribution vanishes in the asymptotic large-system limit, they remain crucial for practically relevant moderate-scale regimes, where efficient learning and robust generalization depend sensitively on the quality of the warm start.

{The numerical experiments further support this theoretical picture. The observed training dynamics of the $20$-qubit system closely follow the predictions of the corresponding linearized kernel model, providing direct empirical evidence for the emergence of a stable lazy-training regime. In this regime, the nonlinear quantum circuit effectively behaves as a fixed kernel method around initialization. The locality-based analysis suggests that the linearized description should become more accurate when ($n \gg L,m$),  provided the relevant kernel spectral quantities remain well controlled. Under this regime, the locality property established in Lemma~\ref{lmma: locality} implies that the light cone of each local observable remains confined to a bounded subset of qubits, which in turn drives the concentration of the quantum neural tangent kernel toward a deterministic limit. For two-dimensional lattices, where the number of Hamiltonian terms grows linearly with the system size, this concentration improves further as the system scales, suggesting that the provably stable linearized trajectory becomes an increasingly accurate description of the optimization dynamics in larger many-body systems. }


Beyond the immediate convergence and generalization guarantees, our results identify the spectrum of the tangent kernel as a key object governing learnability in quantum variational models. This perspective suggests a promising route toward inducing inductive bias in quantum learning architectures through spectral control of the initialized kernel, analogous to recent developments in classical neural tangent kernel theory~\cite{arora2019fine, ronen2019convergence, geifman2023controlling}. From a practical standpoint, the asymptotic equivalence between the quantum model and its linearized counterpart also points toward hybrid algorithmic strategies. In particular, one may envision a quantum-classical pipeline in which a quantum processor is used only to estimate the initialized kernel ($K_{\boldsymbol{\theta}(0)}$), while the downstream learning problem is solved classically using linear methods. Such a strategy may offer a theoretically principled route toward scalable amortized solvers for Hamiltonian prediction tasks and could provide an interesting point of comparison with recent classical approaches that rely on stronger assumptions, such as constant spectral gaps~\cite{huang2022provably}. Understanding whether the kernel induced by guiding-state variational circuits captures features that remain classically inaccessible is therefore an important direction for clarifying the boundary of genuine quantum advantage.

{We also clarify the scope of our generalization guarantee. The bound established in this work should be understood as an in-domain, distributional statement: the training instances \(x_i\) and the unseen test instance \(x'\) are sampled from the same distribution \(\mathcal D\) over the Hamiltonian parameter space, and the guiding-state assumption is required to hold on the support of this distribution. This is the natural regime for amortized prediction, where information from previously solved or measured Hamiltonians is reused to predict properties of statistically related instances. However, the theorem should not be interpreted as an extrapolation guarantee to regions of parameter space where the data distribution has little support, or where the guiding-state oracle fails to provide approximations of the assumed quality. If sufficiently accurate guiding states can only be prepared in a restricted domain, then our guarantee applies only within that domain. Characterizing out-of-distribution behavior and robustness to domain-dependent guiding-state quality remains an important direction for future work.}

{This distributional interpretation also clarifies the role of the guiding-state error parameter \(\delta\). In our main theorem, \(\delta\) is used as a uniform worst-case bound over the support of the data distribution: \( d_{\Tr}(\rho_0(x),\rho(x))\leq \delta \) for every \(x\in\operatorname{supp}(\mathcal D)\). Thus, the guarantee remains valid when the guiding-state quality varies across instances, provided the worst-case error on the relevant distributional domain is bounded by \(\delta\). This worst-case formulation is conservative. In practice, the guiding-state error may be instance-dependent, \(\delta(x)=d_{\Tr}(\rho_0(x),\rho(x))\), and sharper distribution-dependent bounds could potentially be obtained using average-case or high-probability control of \(\delta(x)\). Developing such refinements would require tracking how non-uniform guiding-state errors interact with the empirical tangent-kernel spectrum and the training dynamics, and we leave this for future work.}

Finally, our algorithm relies on the initialization of guiding states and the ALA of $\mathcal{O}(\log(n))$ depth, which generally raises several important challenges. First of all, obtaining the guiding state is not always straightforward. Even though guiding states are known for specific restricted cases~\cite{waite2025physicallymotivatedguidingstateslocal}, the effort required to locate a near-optimal input state in general can be significant—often enough to negate the efficiency gains of our method. This leads to a fundamental question: does the complexity of finding a good initialization merely shift the burden from training a variational quantum algorithm to precomputing an effective starting point? Additionally, there is the question of whether the proposed algorithm contributes to a genuine quantum advantage. If warm-started variational quantum algorithms operate in a parameter regime where classical methods can efficiently approximate their behavior, then their computational benefits may be limited. Recent studies suggest that quantum circuits in well-optimized regions may be classically simulable, raising concerns about whether warm starts truly enable access to classically intractable solutions \cite{drudis2024variational}. 

While our analysis does not provide a definitive resolution to these questions, it highlights an important perspective: the proposed algorithm can serve as a conceptual bridge, motivating the design of quantum-inspired algorithms that leverage guiding state and linearized dynamics. In this way, even if their direct quantum advantage remains uncertain, our framework helps clarify where quantum resources are truly necessary and where classical analogues may suffice.

\section*{Methods}\label{sec: proof_idea}

\subsection*{Notations} We use bold-faced symbols for vectors. For a vector $\boldsymbol{\theta}$, let $\theta_j$ be its $j$-th entry. Similarly, let $A_{ij}$ be the $(i, j)$-th entry of a matrix $A$. Let $\lambda(A)$ be the eigenvalues of a matrix $A$. Let $\textbf{I}$ be the identity matrix and $\mathcal{N}(\boldsymbol{\mu}, \Sigma)$ be the Gaussian distribution with mean $\boldsymbol{\mu}$ and covariance matrix $\Sigma$. 

\subsection*{Guiding States and the Guided Local Hamiltonian Problem}
Our motivation for utilizing guiding states is deeply rooted in the guided local Hamiltonian problem (GLHP) introduced in \cite{gharibian2022dequantizing, gharibian2022improved}. This framework stems from the \textit{local Hamiltonian problem} (LHP), which is the most extensively studied problem in quantum complexity theory. LHP asks whether the ground-state energy of a quantum system, described by a $k$-local Hamiltonian $H =\sum_i H_i$, lies below a threshold $a$ or above a threshold $b$, given a promise gap $b-a \geq 1/\text{poly}(n)$. In this sense, the LHP serves as the canonical abstraction of ground-energy estimation, capturing the essential difficulty of such computations independent of the specific physical model. Kitaev’s seminal result established that the LHP is \textsf{QMA}-complete \cite{kitaev2002classical}, thus making it, in its most general form, intractable even for quantum computers unless $\textsf{BQP} = \textsf{QMA}$.

{While the general LHP is intractable, many systems in physics and chemistry appear to avoid this worst-case behavior. We often leverage classical techniques to find approximate ground states; while these methods sometimes offer formal guarantees, they more frequently rely on heuristics that perform remarkably well in practice, such as DFT in chemistry or mean-field methods in physics. The availability of this "extra" classical information is formalized in the \textit{guided local Hamiltonian problem} (GLHP). In this version, the problem is supplemented with a \emph{guiding state} that provides a starting point with non-trivial overlap with the ground state \cite{gharibian2022dequantizing, gharibian2022improved}. The addition of the guided state to LHP for various physically motivated 2-local Hamiltonians creates a set of problems that \textsf{BQP}-complete\cite{gharibian2022improved}. The \textsf{BQP}-completeness of GLHP establishes that the guiding state assumption, while restricting the problem class, does not trivialize it: the resulting problems remain as hard as anything a quantum computer can efficiently solve.}

{The key computational primitive of GLHP is QPE, because QPE allows extracting the lowest energy and the corresponding eigenstate from the guided state in polynomial time~\cite{cleve1998quantum}. However, despite its algorithmic simplicity and asymptotic efficiency, QPE circuits are notoriously resource-intensive for applications in physics or chemistry~\cite{whitfield2013computational, dalzell2023quantum}.}

{In general, the problem of preparing a guided state for a $k$-local Hamiltonian $H$ is \textsf{QMA}-hard under polynomial-time quantum reductions.  Fortunately, there are many examples in physics and chemistry where guided states can be obtained via efficient classical approximation techniques. In~\cite{waite2025physicallymotivatedguidingstateslocal}, the authors identify a physically motivated guiding states such as fixed-weight states, matrix product states, Gaussian states, and Fendley states and prove \textsf{BQP}-hardness results for them. Beyond these examples, there are numerous techniques that do not have such rigorous guarantees, but have been empirically shown to be remarkably performant for a range of physical systems. }

{In chemistry, Density functional theory (DFT) and Hartree-Fock method can provide very accurate approximations to the ground-state energy and a single Slater determinant state that can be efficiently prepared on a quantum computer~\cite{Chee_2023, Kivlichan_2018}. While there is no guarantee that the ground state will have a sufficient overlap with a single Slater determinant, these methods perform exceptionally well. There are also low-complexity, multi-reference states that can serve as guided states~\cite{tubman2018postponingorthogonalitycatastropheefficient}. Similarly, in condensed matter physics, highly accurate numerical variational methods, such as DMRG and tensor networks, can serve as guided states in practical scenarios. }

\subsection*{Variational Quantum Algorithm}
Variational Quantum Algorithms (VQAs) \cite{cerezo2021variational} are among the most promising strategies for exploring the practical applications of near-term quantum devices. They typically use a hybrid architecture, where quantum computers evaluate the problem's cost function, and a classical algorithm optimizes the parameters of the model. Generally, these algorithms define a parameterized quantum circuit (or a quantum ansatz) $U(\boldsymbol{\theta})$ to generate the output state $\rho(\boldsymbol{\theta})$ from an initialized state $\rho_0$. The core idea is to minimize a loss function $L(\boldsymbol{\theta})$, which encodes the problem we want to solve. Then classical optimization techniques are employed to iteratively adjust the parameters $\boldsymbol{\theta}$, aiming to find the optimal parameters $\boldsymbol{\theta}^{*}$ that give the lowest loss value. The process is illustrated in Figure~\ref{fig: problem_fig}.

One common instance of the optimization algorithm is the gradient descent method, which is intensively used to train classical neural networks \cite{du2019gradient}. Here, the parameters are updated toward the direction of the steepest descent of
the loss function:
\begin{equation}
    \boldsymbol{\theta}(t+1) = \boldsymbol{\theta}(t) - \eta \nabla_{\boldsymbol{\theta}}L(\boldsymbol{\theta})
    \label{eqn: parameter_grad}
\end{equation}
where $\nabla_{\boldsymbol{\theta}}L(\boldsymbol{\theta})$ is the partial gradient vector, $\eta$ is the learning rate parameter controlling the magnitude of the update, and $t$ index the iteration step. The process continues iteratively until the convergence criteria are met. To calculate the partial derivatives $\nabla_{\boldsymbol{\theta}}L(\boldsymbol{\theta})$, one widely used technique in VQAs is the parameter shift rule \cite{mitarai2018quantum}.

Similar to classical neural networks, variational quantum algorithms are largely heuristic in nature. A notable limitation of general VQAs is the challenge of establishing rigorous performance guarantees. Although various studies have been conducted to understand the learning behaviors in training VQAs \cite{you2022convergence, ge2022optimization, anschuetz2022beyond, larocca2023theory}, there are only a few results providing a theoretical performance analysis \cite{farhi2014quantum, schatzki2024theoretical, liu2021rigorous, caro2023out}. Thus, there remains a substantial gap in theoretical analysis in the studies of VQAs. To address this challenge, we would rather shift focus from parameter spaces to function spaces when understanding learning algorithms. Here, the learning algorithms aim to approximate the target function from a pre-determined hypothesis space. The goal of the learning process is thus to find the best approximation within this space. The critical point is defining a suitable hypothesis space for the problem. In classical approaches, the hypothesis space is built up from typical polynomials (polynomial regression) or composition of linear and non-linear functions (neural network), which are supported by rigorous results that these hypothesis spaces are dense in function space \cite{guliyev2018approximation, mhaskar2000fundamentals}. Meanwhile, the quantum hypothesis space is determined by the various choices of ansatzes. Several studies endeavor to manipulate different quantum ansatzes on function space; for example, quantum signal processing (QSP) \cite{motlagh2024generalizedquantumsignalprocessing} provides an ansatz construction to represent any polynomial approximation of the desired function of a unitary. Interestingly, Maria Schuld \cite{schuld2021supervisedquantummachinelearning} noted that many quantum-supervised learning algorithms are fundamentally kernel methods that correspond to reproducing kernel Hilbert spaces (RKHSs). Thus, learning quantum ansatzes is equivalent to learning the `quantum kernel', which generates an RKHS in which the target function lies. Inspired by these results, we aim to explore a mathematical framework for rigorous performance analysis of VQAs.

\subsection*{Neural Tangent Kernel}
One particular framework we employ in this work is the neural tangent kernel (NTK) \cite{jacot2020neuraltangentkernelconvergence, shirai2022quantumtangentkernel}. Let $f_{\boldsymbol{\theta}}$ be the model function defined by the learning algorithm. This function is in a function space $\mathcal{F}$ with respect to a loss function. Without loss of generality, we assume the learning algorithm is defined on mean square error with respect to a training dataset of $\mathcal{S} = \{(x_i, y_i)\}_{i=1}^{M}$:
\begin{equation}
    L_{\mathcal{S}}(\boldsymbol{\theta}) = \frac{1}{2M}\sum_{i=1}^{M} (f_{\boldsymbol{\theta}}(x_i) - y_i)^2.
    \label{eqn: loss_function}
\end{equation}
{Via the gradient flow, we can derive the dynamics of model parameters:
\begin{align}
    \frac{d}{ds}\boldsymbol{\theta}(s) &= -\nabla_{\boldsymbol{\theta}}L_{\mathcal{S}}(\boldsymbol{\theta})|_{\boldsymbol{\theta}(s)}, \quad \forall s>0. \label{eqn: gradient_flow}
\end{align}
To connect this notation with the discrete update in Algorithm~\ref{alg: main_alg}, we work in the small-step-size regime and
 consider the time variable \(t\) through \(s=\eta t\). Equivalently, parameterizing the continuous trajectory by $s = \eta t$, we obtain:
\begin{equation}
    \frac{d}{dt}\boldsymbol{\theta}(t)
    =
    -\eta
    \nabla_{\boldsymbol{\theta}}
    L_{\mathcal{S}}(\boldsymbol{\theta})
    \big|_{\boldsymbol{\theta}=\boldsymbol{\theta}(t)} .
    \label{eqn: iteration_time_gradient_flow}
\end{equation}
For the mean-squared loss, this becomes
\begin{equation}
    \frac{d}{dt}\boldsymbol{\theta}(t)
    =
    -
    \frac{\eta}{M}
    \sum_{i=1}^{M}
    \bigl(f_{\boldsymbol{\theta}(t)}(x_i)-y_i\bigr)
    \nabla_{\boldsymbol{\theta}}
    f_{\boldsymbol{\theta}}(x_i)
    \big|_{\boldsymbol{\theta}=\boldsymbol{\theta}(t)} .
    \label{eqn: iteration_time_gradient_flow_mse}
\end{equation}
Here, \(t\) is interpreted as a continuous interpolation of the discrete
iteration index. At integer times, the derivative \(d\boldsymbol{\theta}(t)/dt\) approximates the finite difference
\(\boldsymbol{\theta}({t+1})-\boldsymbol{\theta}(t)\). Thus, utilizing this continuous-time convention greatly simplifies the derivation of the NTK and lazy-training estimates, which in turn allows us to establish rigorous convergence and generalization guarantees for the discrete iterates of Algorithm~\ref{alg: main_alg}.}

As a result, the model function evolves according to:
\begin{align}
    \frac{d}{dt} f_{\boldsymbol{\theta}(t)}(x) &= \frac{d}{dt}\boldsymbol{\theta}(t).\nabla_{\boldsymbol{\theta}} f_{\boldsymbol{\theta}}(x)|_{\boldsymbol{\theta}(t)}\\
    &= -\frac{\eta}{M}\sum_{i=1}^{M}(f_{\boldsymbol{\theta}(t)}(x_i)-y_i)\cdot{\nabla_{\boldsymbol{\theta}}f_{\boldsymbol{\theta}}(x_i)|_{\boldsymbol{\theta}(t)}}^\top\cdot\nabla_{\boldsymbol{\theta}} f_{\boldsymbol{\theta}}(x)|_{\boldsymbol{\theta}(t)}.
\end{align}
Thus, we can see that the evolution of the model function is indeed governed by a kernel:
\begin{equation}
    K_{\boldsymbol{\theta}}(x, x') := \frac{1}{M} {\nabla_{\boldsymbol{\theta}}f(\boldsymbol{\theta}, x)|_{\boldsymbol{\theta}}}^\top\cdot\nabla_{\boldsymbol{\theta}}f(\boldsymbol{\theta}, x')|_{\boldsymbol{\theta}} 
    \label{eqn: ntk}
\end{equation}
as such: 
\begin{align}
    \frac{d}{dt} f_{\boldsymbol{\theta}(t)}(x) &= -\eta\sum_{i=1}^{M}(f_{\boldsymbol{\theta}(t)}(x_i)-y_i)\cdot K_{\boldsymbol{\theta}(t)}(x_i, x) \label{eqn: grad_f}
\end{align}
This framework enables us to analyze the algorithm’s behavior through the kernel method. 
In the context of classical neural networks, Jacot \textit{et al.}~\cite{jacot2020neuraltangentkernelconvergence} 
showed that, in the over-parameterized regime, $K_{\boldsymbol{\theta}}$ concentrates around its expectation 
over random initializations of $\boldsymbol{\theta}$. Moreover, during gradient descent, 
$K_{\boldsymbol{\theta}}$ remains nearly invariant, a phenomenon known as the \textit{lazy training} regime. Thus, the optimal values of parameters can be calculated using simple linear estimation. However, due to unitary restriction, this nice property does not hold in the quantum settings~\cite{you2023analyzing}. Generally, the quantum neural tangent kernel $K_{\boldsymbol{\theta}}$ still depends on the parameter $\boldsymbol{\theta}$ even in the regime of over-parameterization, making it not applicable with the neural tangent kernel framework. Fortunately, this limitation can be addressed by leveraging the results of \cite{abedi2023quantum}, which identify the conditions, such as a restricted ansatz class and initial loss conditions, under which the model enters the \textit{lazy training} regime. Building on this foundation, we develop a unified performance analysis framework for quantum learning models with guiding states, grounded in kernel theory \cite{shawe2004kernel} and statistical learning theory \cite{shalev2014understanding}.

\subsection*{Concentration of Quantum Neural Tangent Kernel} \label{sec: infinite_width}

The central component of our main results is the concentration property of the quantum neural tangent kernel with ALA, which will be fully described in Supplementary Notes 1 and 2. This result enables us to approximate the model dynamics using a \textit{linearization trick}: a first-order Taylor expansion around the parameter initialization. 

We show that if $U(\boldsymbol{\theta})$ is alternating layered ansatz (Definition~\ref{def: ALA}) with initialization scheme as~\eqref{eqn: param_init}, $K_{\boldsymbol{\theta}}$ is concentrated into a limiting kernel and stays constant via gradient descent algorithms as the system size $n$ goes to infinity. Shown in Algorithm~\ref{alg: main_alg}, we consider the setting of ALA($n, m, p, L$) with $m, p, L \in \mathcal{O}(\log(n))$. Then, let us first discuss the locality lemma regarding our ALA design.
\begin{lmma}
    \label{lmma: locality} Let a parameterized quantum circuit $U(\boldsymbol{\theta}) \in \text{ALA}(n,m,p, L)$. For any $k$-local observable $O$ such that $m \geq k$, $U^{\dagger}(\boldsymbol{\theta})OU(\boldsymbol{\theta})$ acts non-trivially on $\mathcal{O}(Lm)$ qubits.
\end{lmma}
Since $L$ and $m$ are in $\mathcal{O}(\log(n))$, the model function of a $k$-local observable will act non-trivially on qubits in $\mathcal{O}(\log(n)^2)$-size light cones. This locality property of the model function $f_{\boldsymbol{\theta}}$ will aid in proving the convergence of $K_{\boldsymbol{\theta}}$. Specifically, we can show that for any initialization of $\boldsymbol{\theta}$, the tangent kernel entry $K_{\boldsymbol{\theta}}$ defined by ALA will concentrate around its mean. 

\begin{thrm}[Concentration of initialization]
    Consider Algorithm~\ref{alg: main_alg}. {Suppose that the parameters $\boldsymbol{\theta} \in \mathbb{R}^{L \times p \times n/m}$ are drawn from a product distribution; equivalently, the entries of $\boldsymbol{\theta}$ are initialized independently}. Then, the tangent kernel~\eqref{eqn: ntk} satisfies:
     \begin{align}
        \mathbb{P}_{\boldsymbol{\theta}}[\left|K_{\boldsymbol{\theta}}(x, x') - \mathbb{E}_{\boldsymbol{\theta}}[K_{\boldsymbol{\theta}}(x, x')]\right| \geq \epsilon] &\leq \exp{-\Omega\left(\frac{M^2\epsilon^2}{c^4}\cdot \frac{K^4}{n\cdot \text{poly}(\log(n))}\right)}
    \end{align}
    \label{thrm: concentration_paper_version}
\end{thrm}
The theorem shows that the quantum neural tangent kernel is concentrated when the system size $n$ is large, since $K = \mathrm{poly}(n)$. This immediately implies that at the limit of infinite width $n$, the tangent kernel converges to a limit kernel. 

Next, we need to prove that $K_{\boldsymbol{\theta}}$ stays constant during the gradient descent iterations. Leveraging the fact that Theorem~\ref{thrm: concentration_paper_version} is true for any initialization scheme $\boldsymbol{\theta}$. Our initialization scheme in Algorithm~\ref{alg: main_alg} results the following theorem:
\begin{thrm}[Lazy Training]
    Suppose $\boldsymbol{\theta}(0) \sim \mathcal{N}(\boldsymbol{0}, \kappa^2 \mathbf{I})$ with $\kappa \in \Tilde{\mathcal{O}}(\frac{\sqrt{\gamma}\delta}{n})$, then during gradient descent algorithm via loss function~\eqref{eqn: loss_function}, any single entry $K_{ij}(\boldsymbol{\theta})$ of the tangent kernel $K_{\boldsymbol{\theta}}$~\eqref{eqn: ntk} is updated in time by: 
    {\begin{equation}
        |K_{ij}(\boldsymbol{\theta}(t+1)) - K_{ij}(\boldsymbol{\theta}(t))| \leq {\Tilde{\mathcal{O}}\left(  \frac{\eta\delta n}{M\cdot K^3}\right)} \ \ \ \forall i,j
    \end{equation}}
    with probability at least $1 - \gamma$ over the random initialization.
    \label{thrm: lazy_paper_version}
\end{thrm}
Note that we keep $\delta$ in a general form. It could be either a constant or decay with the system size $n$ as in guided local Hamiltonian literature, e.g., $\delta \in \mathcal{O}(1/\text{poly}(n))$. In both cases, Theorem~\ref{thrm: lazy_paper_version} shows that the change of quantum neural tangent kernel after each iteration of the gradient descent algorithm is inverse polynomial in terms of $n$, as $K$ scales polynomially in $n$. This implies the kernel $K_{\boldsymbol{\theta}}$ asymptotically stays constant during training. 

{This concentration phenomenon is reminiscent of the neural tangent kernel (NTK) regime in classical neural networks, yet its origin in our setting is fundamentally different. In classical models, kernel concentration is typically controlled by the \textit{parameter-to-data} ratio. In contrast, here the governing asymptotic parameter is the system size \(n\), and concentration emerges from the \textit{locality structure} of the quantum circuit rather than parameter overparameterization. Specifically, Lemma~\ref{lmma: locality} shows that the expectation of any local observable depends only on the parameters contained in a \textit{bounded light cone} induced by the Alternating Layered Ansatz (ALA). Consequently, the associated tangent kernel concentrates in the regime \(n \gg L\).}

\subsection*{Proofs of Convergence and Generalization} \label{sec: performance_guarantee}
Next, we outline the key ideas behind the proofs of Theorem~\ref{thrm: convergence_paper_ver} and Theorem~\ref{thrm: main_theorem}. The complete technical arguments are provided in Supplementary Notes 3 and 4.

From Theorem~\ref{thrm: concentration_paper_version} and Theorem~\ref{thrm: lazy_paper_version}, we could use results from kernel theory \cite{shawe2004kernel} to analyze the performance of our model. The key to this part is that we simplify the model to its linear approximation when the system size is large enough. To see this, we define a linear model as the first-order Taylor expansion of the model function with respect to the parameters around its initialization:
\begin{equation}
    \hat{f}_{\boldsymbol{\theta}} := f_{\boldsymbol{\theta}(0)} + \nabla_{\boldsymbol{\theta}} f_{\boldsymbol{\theta}}|_{\boldsymbol{\theta}(0)}\cdot(\boldsymbol{\theta} - \boldsymbol{\theta}(0))
\end{equation}
For brevity, we omit the inputs in this analysis. Given $\boldsymbol{\theta}(0)$, $f_{\boldsymbol{\theta}(0)}$ and $\nabla_{\boldsymbol{\theta}} f_{\boldsymbol{\theta}}|_{\boldsymbol{\theta}(0)}$ are all constant, assuming the update step $\eta$ is small enough, {the parameters of the linearized model are updated by gradient descent the same loss function as~\eqref{eqn: loss_function} but for the linear model function $\hat{f}_{\boldsymbol{\theta}}$ 
\begin{equation}
    \hat{L}_{\mathcal{S}}(\boldsymbol{\theta}) := \frac{1}{2M}\sum_{i=1}^{M} (\hat{f}_{\boldsymbol{\theta}}(x_i) - y_i)^2,
    \label{eqn: loss_func_linear}
\end{equation}}
we have:

\begin{align}
    \frac{d}{dt}\boldsymbol{\theta}(t) &= -\eta \nabla_{\boldsymbol{\theta}}\hat{L}_{\mathcal{S}}(\boldsymbol{\theta})|_{\boldsymbol{\theta}(t)} \\
    \frac{d}{dt}\boldsymbol{\theta}(t) &= -\frac{\eta}{M}\sum_{i=1}^{M}(\hat{f}_{\boldsymbol{\theta}(t)}(x_i)-y_i)\cdot \nabla_{\boldsymbol{\theta}} \hat{f}_{\boldsymbol{\theta}}|_{\boldsymbol{\theta}(t)} \\
    \frac{d}{dt}\boldsymbol{\theta}(t) &= -\frac{\eta}{M}\sum_{i=1}^{M}(\hat{f}_{\boldsymbol{\theta}(t)}(x_i)-y_i)\cdot \nabla_{\boldsymbol{\theta}} f_{\boldsymbol{\theta}}|_{\boldsymbol{\theta}(0)} \\
    \frac{d}{dt} \hat{f}_{\boldsymbol{\theta}(t)} &= -\frac{\eta}{M}\sum_{i=1}^{M}(\hat{f}_{\boldsymbol{\theta}(t)}(x_i)-y_i) \cdot {\nabla_{\boldsymbol{\theta}} f_{\boldsymbol{\theta}}|_{\boldsymbol{\theta}(0)}}^\top.\nabla_{\boldsymbol{\theta}} f_{\boldsymbol{\theta}}|_{\boldsymbol{\theta}(0)} \\
    \frac{d}{dt} \hat{f}_{\boldsymbol{\theta}(t)} &= -\eta\sum_{i=1}^{M}(\hat{f}_{\boldsymbol{\theta}(t)}(x_i)-y_i)\cdot K_{\boldsymbol{\theta}(0)} \label{eqn: linearized_model}
\end{align}
Eventually, we recover the same learning dynamics as those of the original model~\eqref{eqn: grad_f}. 
According to Theorems~\ref{thrm: concentration_paper_version} and~\ref{thrm: lazy_paper_version}, 
the kernel \( K_{\boldsymbol{\theta}} \) exhibits strong concentration properties and remains 
effectively constant throughout the gradient descent trajectory. Consequently, in the infinite-width 
(or large-system) limit, we have \( K_{\boldsymbol{\theta}(0)} = K_{\boldsymbol{\theta}(t)} \) for all \( t \), 
indicating that the training dynamics are governed by a fixed kernel. This observation implies that, 
as the system size approaches infinity, the true nonlinear model converges to its linearized counterpart, 
validating the linear approximation regime. 

We now proceed to analyze the convergence behavior of this linearized model~\eqref{eqn: linearized_model}, as formalized in the following theorem.

\begin{thrm}\label{thrm: convergence_of_linear_model_paper_version}
    Consider Algorithm~\ref{alg: main_alg}, and let the eigenvalues of the kernel matrix 
\( K_{\boldsymbol{\theta}(0)} \) satisfy
{\(
0 < \lambda_{\min} := \lambda_{\min}(K_{\boldsymbol{\theta}(0)}) 
\leq \lambda_j 
\leq \lambda_{\max} := \lambda_{\max}(K_{\boldsymbol{\theta}(0)}) < \infty.
\)}
For a learning rate $\eta \leq \frac{1}{\lambda_{\max}}$ and parameter 
\( \kappa \in \tilde{\mathcal{O}}\!\left({\sqrt{\gamma}\,\delta/n}\right) \), the loss \( \hat{L}_{\mathcal{S}}(\boldsymbol{\theta}(t)) \)~\eqref{eqn: loss_func_linear} at time \( t \) satisfies
\begin{equation}
    |\hat{L}_{\mathcal{S}}(\boldsymbol{\theta}(t))| 
    \leq 
    \tilde{\mathcal{O}}\!\left(
        \sum_{j}
        (1 - \eta \lambda_j)^{2t}\, \delta^2
    \right).
\end{equation}
\end{thrm}

However, the true model coincides with its linear approximation only in the limit of infinite system size. Our interest lies instead in the practically relevant case of finite $n$. We therefore establish the following training error bound between the true model and its linear approximation:
\begin{thrm}\label{thrm: training_loss_bound_paper_version}
    Consider the Algorithm~\ref{alg: main_alg} trained on a dataset $\mathcal{S}$~\eqref{eqn: training_dataset} and loss function~\eqref{eqn: loss_function}. Suppose $\eta \leq \frac{1}{\lambda_{\max}}$, $\kappa \in \Tilde{\mathcal{O}}(\frac{\sqrt{\gamma}\delta}{n})$, we have:
    \begin{equation}
        \left| L_{\mathcal{S}}(\boldsymbol{\theta}(t)) - \hat{L}_{\mathcal{S}}(\boldsymbol{\theta}(t))\right| \leq \Tilde{\mathcal{O}}\left(\frac{n}{K^3}\eta^2 t^2 \delta^{3}\right)
    \end{equation}
\end{thrm}
Combining Theorem~\ref{thrm: convergence_of_linear_model_paper_version} and Theorem~\ref{thrm: training_loss_bound_paper_version}, we derive the convergence result in Theorem~\ref{thrm: convergence_paper_ver}.

The result in Theorem~\ref{thrm: training_loss_bound_paper_version} will also help us derive the theoretical bound of the generalization error of the true model. Particularly, we will provide generalization bounds based on the Rademacher complexity \cite{shalev2014understanding}. Generally, Rademacher complexity is a measure used in statistical learning theory to quantify the ability of a class of functions to fit random noise. It evaluates the richness of a function class by calculating the average correlation between the functions and random Rademacher variables, which take values of +1 or -1 with equal probability. Lower Rademacher complexity indicates a less flexible model, typically leading to better generalization from training data to unseen data. Due to the concentration of tangent kernel at the infinite-width limit, the class of functions of the model is well-characterized by famous kernel methods \cite{shawe2004kernel}. 

Let us define $\mathcal{F}$ to be the set of functions that could be learned from a learning model. The Rademacher complexity theory allows us to obtain bounds on the generalization error associated with learning from training data \cite{shalev2014understanding}. For convenience, we denote $\ell(f_{\boldsymbol{\theta}}(x)) := \left|f_{\boldsymbol{\theta}
}(x) - \Tr[O\rho(x)]\right|$, the Rademacher complexity of a function space $\ell \circ \mathcal{F}$ with respect to training data $\mathcal{S}$ is defined as follows:
\begin{equation}
    R(\ell \circ \mathcal{F} \circ \mathcal{S}) := \frac{1}{M} \mathbb{E}_{\boldsymbol{\sigma} \in \{-1, 1\}^{M}} \left[\sup \sum_{i=1}^{M} \sigma_i \ell(f_{\boldsymbol{\theta}(t)}(x_i))\right]
\end{equation}
This quantity provides a bound on the generalization error by the following lemma
\begin{lmma}[Theorem 26.5 \cite{shalev2014understanding}]
    For a training sample $\mathcal{S} = \{x_1,\dots, x_M\}$ generated by an unknown distribution $\mathcal{D}$ and real-value function class $\mathcal{F}$, such that for all $x$ and $f\in \mathcal{F}$ we have $|\ell(f(x))| \leq c$. Then, for a confidence parameter $\gamma \in (0,1)$, with probability at least $1-\gamma$ over the choice of $\mathcal{S}$, every $f\in \mathcal{F}$ satisfies:
    \begin{equation}
        \mathbb{E}_{x\sim\mathcal{D}}[\ell(f(x))] - \mathbb{E}_{x\sim\mathcal{S}}[\ell(f(x))] \leq 2 R(\ell \circ \mathcal{F} \circ \mathcal{S}) + 4c\sqrt{\frac{2\ln(4/\gamma)}{M}}
    \end{equation}
    \label{lmma: generalization_error_paper_version}
\end{lmma}
The lemma shows that the generalization error is upper-bounded by the Rademacher complexity. If the quantity $R(\ell \circ \mathcal{F} \circ \mathcal{S})$ is small, then we could reliably learn the target function. Thus, we next aim to bound this quantity. 

First, we denote the Rademacher complexity of the function class, $\hat{\mathcal{F}}$, as the set of functions generated by the initialization kernel $K_{\boldsymbol{\theta}(0)}$ based on the dynamics in~\eqref{eqn: linearized_model}. As pointed out in \cite{jacot2020neuraltangentkernelconvergence} and our result in Theorem~\ref{thrm: convergence_paper_ver}, we see that the convergence of the tangent kernel model is faster along the eigenspaces with larger eigenvalues $\lambda_i$ of $K_{\boldsymbol{\theta}(0)}$. We are typically interested in the case where the model focuses on fitting the most relevant kernel principal components (larger eigenvalues), which is the motivation for the use of early stopping. Thus, for the analysis of Rademacher complexity, we consider the function class with a bounded sum of eigenvalues:
\begin{equation}
    (\ell\circ\mathcal{F})_B := \{\ell_t \in \ell\circ {\mathcal{F}} | B_1\leq \sum_{i}(1-\eta \lambda_i)^t \leq B_2\}
\end{equation}
From that, we can bound the Rademacher complexity to the function class of the linear model in~\eqref{eqn: linearized_model} as:
\begin{equation}
    R((\ell\circ\hat{\mathcal{F}})_B\circ\mathcal{S}) \leq  B_2\sqrt{\frac{2}{M}L_{\mathcal{S}}(\boldsymbol{\theta}(0))}.
    \label{temp: 03}
\end{equation}
Finally, we compare the Rademacher complexity of the true model with its linear approximation as such:
\begin{equation}
    {|R((\ell\circ\mathcal{F})_B\circ \mathcal{S}) - R((\ell\circ\hat{\mathcal{F}})_B\circ \mathcal{S})| \leq \Tilde{\mathcal{O}}\left(\frac{n}{K^3} \left(\frac{\ln(M/B_1)}{\lambda_{\min}}\right)^2 \delta^2\right)} 
    \label{temp: 08}
\end{equation}
Combining the results from~\eqref{temp: 03} and~\eqref{temp: 08}, we establish the bound in Theorem~\ref{thrm: main_theorem}. The detailed proofs for~\eqref{temp: 03} and~\eqref{temp: 08} are presented in Supplementary Note 4. 




\section*{Declarations}

\paragraph{Data Availability.} All data used in this study are synthetically generated. The code required to reproduce the datasets and numerical experiments is publicly available in the repository listed under Code availability.

\paragraph{Code availability.} The custom code used to implement the guiding-state variational quantum algorithm, generate Hamiltonian instances, and reproduce the numerical experiments in this study is publicly available in the GitHub repository \texttt{analytic\_qnn\_GLH}, accessible at \url{https://github.com/ichirokira/analytic_qnn_GLH}. The repository includes the training scripts, model implementations, configuration files, and visualization notebooks used to generate the results reported in the main text and Supplementary Information.

\paragraph{Acknowledgments.} We thank Thinh Le for the insightful discussion on the generalization error of kernel-based models. We also thank Marco Cerezo for the recommendation of hardness in preparing guiding states. TN is supported by a scholarship from the Sydney Quantum Academy, PHDR06031 and the ARC Centre of Excellence for Quantum Computation and Communication Technology (CQC2T), project number CE170100012.  MK acknowledges the ARC Centre of Excellence for Quantum Computation and
Communication Technology (CQC2T), project number CE170100012.

\paragraph{Author contributions.} M.K. initiated the research direction and supervised the project. T.N. extended the original idea, designed the variational algorithm, developed the theoretical framework, proofs, and numerical experiments. T.N. also drafted the manuscript. Both authors reviewed and approved the final version.

\paragraph{Competing Interests.} The authors declare no competing financial or non-financial interests.

\medskip 

\printbibliography

@inproceedings{gharibian2022dequantizing,
  title={Dequantizing the quantum singular value transformation: hardness and applications to quantum chemistry and the quantum PCP conjecture},
  author={Gharibian, Sevag and Le Gall, Fran{\c{c}}ois},
  booktitle={Proceedings of the 54th Annual ACM SIGACT Symposium on Theory of Computing},
  pages={19--32},
  year={2022}
}

@book{kitaev2002classical,
  title={Classical and quantum computation},
  author={Kitaev, Alexei Yu and Shen, Alexander and Vyalyi, Mikhail N},
  number={47},
  year={2002},
  publisher={American Mathematical Soc.}
}

@article{gharibian2022improved,
  title={Improved hardness results for the guided local hamiltonian problem},
  author={Gharibian, Sevag and Hayakawa, Ryu and Gall, Fran{\c{c}}ois Le and Morimae, Tomoyuki},
  journal={arXiv preprint arXiv:2207.10250},
  
  year={2022}
}

@misc{waite2025physicallymotivatedguidingstateslocal,
  title={Physically-Motivated Guiding States for Local Hamiltonians}, 
  author={Gabriel Waite and Karl Lin and Samuel J Elman and Michael J Bremner},
  year={2025},
  eprint={2509.25815},
  archivePrefix={arXiv},
  primaryClass={quant-ph},
  url={https://arxiv.org/abs/2509.25815}, 
}

@article{huang2022provably,
  title={Provably efficient machine learning for quantum many-body problems},
  author={Huang, Hsin-Yuan and Kueng, Richard and Torlai, Giacomo and Albert, Victor V and Preskill, John},
  journal={Science},
  journal={Science},
  volume={377},
  number={6613},
  pages={eabk3333},
  year={2022},
  publisher={American Association for the Advancement of Science}
}

@article{cerezo2021variational,
  title={Variational quantum algorithms},
  author={Cerezo, Marco and Arrasmith, Andrew and Babbush, Ryan and Benjamin, Simon C and Endo, Suguru and Fujii, Keisuke and McClean, Jarrod R and Mitarai, Kosuke and Yuan, Xiao and Cincio, Lukasz and others},
  journal={Nature Reviews Physics},
  journal={Nat. Rev. Phys.},
  volume={3},
  number={9},
  pages={625--644},
  year={2021},
  publisher={Nature Publishing Group UK London}
}

@misc{jacot2020neuraltangentkernelconvergence,
  title={Neural Tangent Kernel: Convergence and Generalization in Neural Networks}, 
  author={Arthur Jacot and Franck Gabriel and Clément Hongler},
  year={2020},
  eprint={1806.07572},
  archivePrefix={arXiv},
  primaryClass={cs.LG},
  url={https://arxiv.org/abs/1806.07572}, 
}

@article{liu2023analytic,
  title={Analytic theory for the dynamics of wide quantum neural networks},
  author={Liu, Junyu and Najafi, Khadijeh and Sharma, Kunal and Tacchino, Francesco and Jiang, Liang and Mezzacapo, Antonio},
  journal={Physical Review Letters},
  journal={Phys. Rev. Lett.},
  volume={130},
  number={15},
  pages={150601},
  year={2023},
  publisher={APS}
}

@article{abedi2023quantum,
  title={Quantum lazy training},
  author={Abedi, Erfan and Beigi, Salman and Taghavi, Leila},
  journal={Quantum},
  journal={Quantum},
  volume={7},
  pages={989},
  year={2023},
  publisher={Verein zur F{\"o}rderung des Open Access Publizierens in den Quantenwissenschaften}
}

@article{kandala2017hardware,
  title={Hardware-efficient variational quantum eigensolver for small molecules and quantum magnets},
  author={Kandala, Abhinav and Mezzacapo, Antonio and Temme, Kristan and Takita, Maika and Brink, Markus and Chow, Jerry M and Gambetta, Jay M},
  journal={nature},
  journal={Nature},
  volume={549},
  number={7671},
  pages={242--246},
  year={2017},
  publisher={Nature Publishing Group}
}

@article{mcclean2018barren,
  title={Barren plateaus in quantum neural network training landscapes},
  author={McClean, Jarrod R and Boixo, Sergio and Smelyanskiy, Vadim N and Babbush, Ryan and Neven, Hartmut},
  journal={Nature communications},
  journal={Nat. Commun.},
  volume={9},
  number={1},
  pages={4812},
  year={2018},
  publisher={Nature Publishing Group UK London}
}

@article{zhang2022escaping,
  title={Escaping from the barren plateau via gaussian initializations in deep variational quantum circuits},
  author={Zhang, Kaining and Liu, Liu and Hsieh, Min-Hsiu and Tao, Dacheng},
  journal={Advances in Neural Information Processing Systems},
  journal={Adv. Neural Inf. Process. Syst.},
  volume={35},
  pages={18612--18627},
  year={2022}
}

@article{cerezo2021cost,
  title={Cost function dependent barren plateaus in shallow parametrized quantum circuits},
  author={Cerezo, Marco and Sone, Akira and Volkoff, Tyler and Cincio, Lukasz and Coles, Patrick J},
  journal={Nature communications},
  journal={Nat. Commun.},
  volume={12},
  number={1},
  pages={1791},
  year={2021},
  publisher={Nature Publishing Group UK London}
}

@article{nakaji2021expressibility,
  title={Expressibility of the alternating layered ansatz for quantum computation},
  author={Nakaji, Kouhei and Yamamoto, Naoki},
  journal={Quantum},
  journal={Quantum},
  volume={5},
  pages={434},
  year={2021},
  publisher={Verein zur F{\"o}rderung des Open Access Publizierens in den Quantenwissenschaften}
}

@article{farhi2014quantum,
  title={A quantum approximate optimization algorithm},
  author={Farhi, Edward and Goldstone, Jeffrey and Gutmann, Sam},
  journal={arXiv preprint arXiv:1411.4028},
  
  year={2014}
}

@article{schatzki2024theoretical,
  title={Theoretical guarantees for permutation-equivariant quantum neural networks},
  author={Schatzki, Louis and Larocca, Martin and Nguyen, Quynh T and Sauvage, Frederic and Cerezo, Marco},
  journal={npj Quantum Information},
  journal={npj Quantum Inf.},
  volume={10},
  number={1},
  pages={12},
  year={2024},
  publisher={Nature Publishing Group UK London}
}

@article{brandao2016local,
  title={Local random quantum circuits are approximate polynomial-designs},
  author={Brandao, Fernando GSL and Harrow, Aram W and Horodecki, Micha{\l}},
  journal={Communications in Mathematical Physics},
  journal={Commun. Math. Phys.},
  volume={346},
  pages={397--434},
  year={2016},
  publisher={Springer}
}

@article{harrow2023approximate,
  title={Approximate unitary t-designs by short random quantum circuits using nearest-neighbor and long-range gates},
  author={Harrow, Aram W and Mehraban, Saeed},
  journal={Communications in Mathematical Physics},
  journal={Commun. Math. Phys.},
  volume={401},
  number={2},
  pages={1531--1626},
  year={2023},
  publisher={Springer}
}

@book{shalev2014understanding,
  title={Understanding machine learning: From theory to algorithms},
  author={Shalev-Shwartz, Shai and Ben-David, Shai},
  year={2014},
  publisher={Cambridge university press}
}

@book{shawe2004kernel,
  title={Kernel methods for pattern analysis},
  author={Shawe-Taylor, John and Cristianini, Nello},
  year={2004},
  publisher={Cambridge university press}
}

@article{cleve1998quantum,
  title={Quantum algorithms revisited},
  author={Cleve, Richard and Ekert, Artur and Macchiavello, Chiara and Mosca, Michele},
  journal={Proceedings of the Royal Society of London. Series A: Mathematical, Physical and Engineering Sciences},
  journal={Proc. R. Soc. A},
  volume={454},
  number={1969},
  pages={339--354},
  year={1998},
  publisher={The Royal Society}
}

@article{dalzell2023quantum,
  title={Quantum algorithms: A survey of applications and end-to-end complexities},
  author={Dalzell, Alexander M and McArdle, Sam and Berta, Mario and Bienias, Przemyslaw and Chen, Chi-Fang and Gily{\'e}n, Andr{\'a}s and Hann, Connor T and Kastoryano, Michael J and Khabiboulline, Emil T and Kubica, Aleksander and others},
  journal={arXiv preprint arXiv:2310.03011},
  
  year={2023}
}

@article{guliyev2018approximation,
  title={Approximation capability of two hidden layer feedforward neural networks with fixed weights},
  author={Guliyev, Namig J and Ismailov, Vugar E},
  journal={Neurocomputing},
  journal={Neurocomputing},
  volume={316},
  pages={262--269},
  year={2018},
  publisher={Elsevier}
}

@book{mhaskar2000fundamentals,
  title={Fundamentals of approximation theory},
  author={Mhaskar, Hrushikesh Narhar and Pai, Devidas V},
  year={2000},
  publisher={CRC Press}
}

@misc{motlagh2024generalizedquantumsignalprocessing,
  title={Generalized Quantum Signal Processing}, 
  author={Danial Motlagh and Nathan Wiebe},
  year={2024},
  eprint={2308.01501},
  archivePrefix={arXiv},
  primaryClass={quant-ph},
  url={https://arxiv.org/abs/2308.01501}, 
}

@misc{schuld2021supervisedquantummachinelearning,
  title={Supervised quantum machine learning models are kernel methods}, 
  author={Maria Schuld},
  year={2021},
  eprint={2101.11020},
  archivePrefix={arXiv},
  primaryClass={quant-ph},
  url={https://arxiv.org/abs/2101.11020}, 
}

@article{kubler2021inductive,
  title={The inductive bias of quantum kernels},
  author={K{\"u}bler, Jonas and Buchholz, Simon and Sch{\"o}lkopf, Bernhard},
  journal={Advances in Neural Information Processing Systems},
  journal={Adv. Neural Inf. Process. Syst.},
  volume={34},
  pages={12661--12673},
  year={2021}
}

@article{liu2021rigorous,
  title={A rigorous and robust quantum speed-up in supervised machine learning},
  author={Liu, Yunchao and Arunachalam, Srinivasan and Temme, Kristan},
  journal={Nature Physics},
  journal={Nat. Phys.},
  volume={17},
  number={9},
  pages={1013--1017},
  year={2021},
  publisher={Nature Publishing Group UK London}
}

@article{whitfield2013computational,
  title={Computational complexity in electronic structure},
  author={Whitfield, James Daniel and Love, Peter John and Aspuru-Guzik, Al{\'a}n},
  journal={Physical Chemistry Chemical Physics},
  journal={Phys. Chem. Chem. Phys.},
  volume={15},
  number={2},
  pages={397--411},
  year={2013},
  publisher={Royal Society of Chemistry}
}

@article{mitarai2018quantum,
  title={Quantum circuit learning},
  author={Mitarai, Kosuke and Negoro, Makoto heart and Kitagawa, Masahiro and Fujii, Keisuke},
  journal={Physical Review A},
  journal={Phys. Rev. A},
  volume={98},
  number={3},
  pages={032309},
  year={2018},
  publisher={APS}
}

@inproceedings{du2019gradient,
  title={Gradient descent finds global minima of deep neural networks},
  author={Du, Simon and Lee, Jason and Li, Haochuan and Wang, Liwei and Zhai, Xiyu},
  booktitle={International conference on machine learning},
  pages={1675--1685},
  year={2019},
  organization={PMLR}
}

@article{you2022convergence,
  title={A convergence theory for over-parameterized variational quantum eigensolvers},
  author={You, Xuchen and Chakrabarti, Shouvanik and Wu, Xiaodi},
  journal={arXiv preprint arXiv:2205.12481},
  
  year={2022}
}

@article{ge2022optimization,
  title={The optimization landscape of hybrid quantum--classical algorithms: From quantum control to NISQ applications},
  author={Ge, Xiaozhen and Wu, Re-Bing and Rabitz, Herschel},
  journal={Annual Reviews in Control},
  journal={Annu. Rev. Control},
  volume={54},
  pages={314--323},
  year={2022},
  publisher={Elsevier}
}

@article{anschuetz2022beyond,
  title={Beyond barren plateaus: Quantum variational algorithms are swamped with traps. 2022},
  author={Anschuetz, Eric R and Kiani, Bobak T},
  journal={arXiv preprint arXiv:2205.05786},
  
  year={2022}
}

@article{larocca2023theory,
  title={Theory of overparametrization in quantum neural networks},
  author={Larocca, Martin and Ju, Nathan and Garc{\'\i}a-Mart{\'\i}n, Diego and Coles, Patrick J and Cerezo, Marco},
  journal={Nature Computational Science},
  journal={Nat. Comput. Sci.},
  volume={3},
  number={6},
  pages={542--551},
  year={2023},
  publisher={Nature Publishing Group US New York}
}

@article{caro2023out,
  title={Out-of-distribution generalization for learning quantum dynamics},
  author={Caro, Matthias C and Huang, Hsin-Yuan and Ezzell, Nicholas and Gibbs, Joe and Sornborger, Andrew T and Cincio, Lukasz and Coles, Patrick J and Holmes, Zo{\"e}},
  journal={Nature Communications},
  journal={Nat. Commun.},
  volume={14},
  number={1},
  pages={3751},
  year={2023},
  publisher={Nature Publishing Group UK London}
}

@misc{shirai2022quantumtangentkernel,
  title={Quantum tangent kernel}, 
  author={Norihito Shirai and Kenji Kubo and Kosuke Mitarai and Keisuke Fujii},
  year={2022},
  eprint={2111.02951},
  archivePrefix={arXiv},
  primaryClass={quant-ph},
  url={https://arxiv.org/abs/2111.02951}, 
}

@article{peruzzo2014variational,
  title={A variational eigenvalue solver on a photonic quantum processor},
  author={Peruzzo, Alberto and McClean, Jarrod and Shadbolt, Peter and Yung, Man-Hong and Zhou, Xiao-Qi and Love, Peter J and Aspuru-Guzik, Al{\'a}n and O’brien, Jeremy L},
  journal={Nature communications},
  journal={Nat. Commun.},
  volume={5},
  number={1},
  pages={4213},
  year={2014},
  publisher={Nature Publishing Group UK London}
}

@inproceedings{arora2019fine,
  title={Fine-grained analysis of optimization and generalization for overparameterized two-layer neural networks},
  author={Arora, Sanjeev and Du, Simon and Hu, Wei and Li, Zhiyuan and Wang, Ruosong},
  booktitle={International Conference on Machine Learning},
  pages={322--332},
  year={2019},
  organization={PMLR}
}

@article{ronen2019convergence,
  title={The convergence rate of neural networks for learned functions of different frequencies},
  author={Ronen, Basri and Jacobs, David and Kasten, Yoni and Kritchman, Shira},
  journal={Advances in Neural Information Processing Systems},
  journal={Adv. Neural Inf. Process. Syst.},
  volume={32},
  year={2019}
}

@article{geifman2023controlling,
  title={Controlling the Inductive Bias of Wide Neural Networks by Modifying the Kernel's Spectrum},
  author={Geifman, Amnon and Barzilai, Daniel and Basri, Ronen and Galun, Meirav},
  journal={arXiv preprint arXiv:2307.14531},
  
  year={2023}
}

@article{Ragone_2024,
  title={A Lie algebraic theory of barren plateaus for deep parameterized quantum circuits},
  volume={15},
  ISSN={2041-1723},
  url={http://dx.doi.org/10.1038/s41467-024-49909-3},
  DOI={10.1038/s41467-024-49909-3},
  number={1},
  journal={Nature Communications},
  journal={Nat. Commun.},
  publisher={Springer Science and Business Media LLC},
  author={Ragone, Michael and Bakalov, Bojko N. and Sauvage, Frédéric and Kemper, Alexander F. and Ortiz Marrero, Carlos and Larocca, Martín and Cerezo, M.},
  year={2024},
  month=aug 
}

@misc{marrero2021entanglementinducedbarrenplateaus,
  title={Entanglement Induced Barren Plateaus}, 
  author={Carlos Ortiz Marrero and Mária Kieferová and Nathan Wiebe},
  year={2021},
  eprint={2010.15968},
  archivePrefix={arXiv},
  primaryClass={quant-ph},
  url={https://arxiv.org/abs/2010.15968}, 
}

@article{drudis2024variational,
  title={Variational quantum simulation: a case study for understanding warm starts},
  author={Drudis, Marc and Thanasilp, Supanut and Holmes, Zo{\"e} and others},
  journal={arXiv preprint arXiv:2404.10044},
  
  year={2024}
}

@misc{meyer2024warmstartvariationalquantumpolicy,
  title={Warm-Start Variational Quantum Policy Iteration}, 
  author={Nico Meyer and Jakob Murauer and Alexander Popov and Christian Ufrecht and Axel Plinge and Christopher Mutschler and Daniel D. Scherer},
  year={2024},
  eprint={2404.10546},
  archivePrefix={arXiv},
  primaryClass={quant-ph},
  url={https://arxiv.org/abs/2404.10546}, 
}

@article{egger2021warm,
  title={Warm-starting quantum optimization},
  author={Egger, Daniel J and Mare{\v{c}}ek, Jakub and Woerner, Stefan},
  journal={Quantum},
  journal={Quantum},
  volume={5},
  pages={479},
  year={2021},
  publisher={Verein zur F{\"o}rderung des Open Access Publizierens in den Quantenwissenschaften}
}

@article{park2024hamiltonian,
  title={Hamiltonian variational ansatz without barren plateaus},
  author={Park, Chae-Yeun and Killoran, Nathan},
  journal={Quantum},
  journal={Quantum},
  volume={8},
  pages={1239},
  year={2024},
  publisher={Verein zur F{\"o}rderung des Open Access Publizierens in den Quantenwissenschaften}
}

@article{park2024hardware,
  title={Hardware-efficient ansatz without barren plateaus in any depth},
  author={Park, Chae-Yeun and Kang, Minhyeok and Huh, Joonsuk},
  journal={arXiv preprint arXiv:2403.04844},
  
  year={2024}
}

@misc{shi2024avoidingbarrenplateausgaussian,
  title={Avoiding barren plateaus via Gaussian Mixture Model}, 
  author={Xiao Shi and Yun Shang},
  year={2024},
  eprint={2402.13501},
  archivePrefix={arXiv},
  primaryClass={quant-ph},
  url={https://arxiv.org/abs/2402.13501}, 
}

@article{brandao2018fixed,
  title={For fixed control parameters the quantum approximate optimization algorithm's objective function value concentrates for typical instances},
  author={Brandao, Fernando GSL and Broughton, Michael and Farhi, Edward and Gutmann, Sam and Neven, Hartmut},
  journal={arXiv preprint arXiv:1812.04170},
  
  year={2018}
}

@article{farhi2022quantum,
  title={The quantum approximate optimization algorithm and the Sherrington-Kirkpatrick model at infinite size},
  author={Farhi, Edward and Goldstone, Jeffrey and Gutmann, Sam and Zhou, Leo},
  journal={Quantum},
  journal={Quantum},
  volume={6},
  pages={759},
  year={2022},
  publisher={Verein zur F{\"o}rderung des Open Access Publizierens in den Quantenwissenschaften}
}

@misc{lewis2023improvedmachinelearningalgorithm,
  title={Improved machine learning algorithm for predicting ground state properties}, 
  author={Laura Lewis and Hsin-Yuan Huang and Viet T. Tran and Sebastian Lehner and Richard Kueng and John Preskill},
  year={2023},
  eprint={2301.13169},
  archivePrefix={arXiv},
  primaryClass={quant-ph},
  url={https://arxiv.org/abs/2301.13169}, 
}

@misc{mhiri2025unifyingaccountwarmstart,
  title={A unifying account of warm start guarantees for patches of quantum landscapes}, 
  author={Hela Mhiri and Ricard Puig and Sacha Lerch and Manuel S. Rudolph and Thiparat Chotibut and Supanut Thanasilp and Zoë Holmes},
  year={2025},
  eprint={2502.07889},
  archivePrefix={arXiv},
  primaryClass={quant-ph},
  url={https://arxiv.org/abs/2502.07889}, 
}

@inproceedings{you2023analyzing,
  title={Analyzing convergence in quantum neural networks: deviations from neural tangent kernels},
  author={You, Xuchen and Chakrabarti, Shouvanik and Chen, Boyang and Wu, Xiaodi},
  booktitle={International Conference on Machine Learning},
  pages={40199--40224},
  year={2023},
  organization={PMLR}
}

@article{Carleo_2019,
  title={Machine learning and the physical sciences},
  volume={91},
  ISSN={1539-0756},
  url={http://dx.doi.org/10.1103/RevModPhys.91.045002},
  DOI={10.1103/revmodphys.91.045002},
  number={4},
  journal={Reviews of Modern Physics},
  journal={Rev. Mod. Phys.},
  publisher={American Physical Society (APS)},
  author={Carleo, Giuseppe and Cirac, Ignacio and Cranmer, Kyle and Daudet, Laurent and Schuld, Maria and Tishby, Naftali and Vogt-Maranto, Leslie and Zdeborová, Lenka},
  year={2019},
  month=dec 
}

@article{Carrasquilla_2020,
  title={Machine learning for quantum matter},
  volume={5},
  ISSN={2374-6149},
  url={http://dx.doi.org/10.1080/23746149.2020.1797528},
  DOI={10.1080/23746149.2020.1797528},
  number={1},
  journal={Advances in Physics: X},
  journal={Adv. Phys. X},
  publisher={Informa UK Limited},
  author={Carrasquilla, Juan},
  year={2020},
  month=jan, 
  pages={1797528} 
}

@incollection{weisse2008exact,
  title={Exact diagonalization techniques},
  author={Wei{\ss}e, Alexander and Fehske, Holger},
  booktitle={Computational many-particle physics},
  pages={529--544},
  year={2008},
  publisher={Springer}
}

@article{carleo2017solving,
  title={Solving the quantum many-body problem with artificial neural networks},
  author={Carleo, Giuseppe and Troyer, Matthias},
  journal={Science},
  journal={Science},
  volume={355},
  number={6325},
  pages={602--606},
  year={2017},
  publisher={American Association for the Advancement of Science}
}

@article{white1992density,
  title={Density matrix formulation for quantum renormalization groups},
  author={White, Steven R},
  journal={Physical review letters},
  journal={Phys. Rev. Lett.},
  volume={69},
  number={19},
  pages={2863},
  year={1992},
  publisher={APS}
}

@book{nightingale1998quantum,
  title={Quantum Monte Carlo methods in physics and chemistry},
  author={Nightingale, M Peter and Umrigar, Cyrus J},
  number={525},
  year={1998},
  publisher={Springer Science \& Business Media}
}

@article{Chee_2023,
  title={Shallow quantum circuits for efficient preparation of Slater determinants and correlated states on a quantum computer},
  volume={108},
  ISSN={2469-9934},
  url={http://dx.doi.org/10.1103/PhysRevA.108.022416},
  DOI={10.1103/physreva.108.022416},
  number={2},
  journal={Physical Review A},
  journal={Phys. Rev. A},
  publisher={American Physical Society (APS)},
  author={Chee, Chong Hian and Leykam, Daniel and Mak, Adrian M. and Angelakis, Dimitris G.},
  year={2023},
  month=aug 
}

@misc{belaloui2024groundstateenergyestimation,
  title={Ground State Energy Estimation on Current Quantum Hardware Through The Variational Quantum Eigensolver: A Comprehensive Study}, 
  author={Nacer Eddine Belaloui and Abdellah Tounsi and Rabah Abdelmouheymen Khamadja and Mohamed Messaoud Louamri and Achour Benslama and David E. Bernal Neira and Mohamed Taha Rouabah},
  year={2024},
  eprint={2412.02606},
  archivePrefix={arXiv},
  primaryClass={quant-ph},
  url={https://arxiv.org/abs/2412.02606}, 
}

@inproceedings{Lolur_2021,
  title={Benchmarking the variational quantum eigensolver through simulation of the ground state energy of prebiotic molecules on high-performance computers},
  volume={2362},
  ISSN={0094-243X},
  url={http://dx.doi.org/10.1063/5.0054915},
  DOI={10.1063/5.0054915},
  booktitle={MIPT (PHYSTECH) - QUANT 2020},
  publisher={AIP Publishing},
  author={Lolur, P. and Rahm, M. and Skogh, M. and García-Álvarez, L. and Wendin, G.},
  year={2021},
  pages={030005} 
}

@article{de_Keijzer_2022,
  title={Optimization of the variational quantum eigensolver for quantum chemistry applications},
  volume={4},
  ISSN={2639-0213},
  url={http://dx.doi.org/10.1116/5.0076435},
  DOI={10.1116/5.0076435},
  number={1},
  journal={AVS Quantum Science},
  journal={AVS Quantum Sci.},
  publisher={American Vacuum Society},
  author={de Keijzer, R. J. P. T. and Colussi, V. E. and Škorić, B. and Kokkelmans, S. J. J. M. F.},
  year={2022},
  month=feb 
}

@article{Kivlichan_2018,
  title={Quantum Simulation of Electronic Structure with Linear Depth and Connectivity},
  volume={120},
  ISSN={1079-7114},
  url={http://dx.doi.org/10.1103/PhysRevLett.120.110501},
  DOI={10.1103/physrevlett.120.110501},
  number={11},
  journal={Physical Review Letters},
  journal={Phys. Rev. Lett.},
  publisher={American Physical Society (APS)},
  author={Kivlichan, Ian D. and McClean, Jarrod and Wiebe, Nathan and Gidney, Craig and Aspuru-Guzik, Alán and Chan, Garnet Kin-Lic and Babbush, Ryan},
  year={2018},
  month=mar 
}

@misc{tubman2018postponingorthogonalitycatastropheefficient,
  title={Postponing the orthogonality catastrophe: efficient state preparation for electronic structure simulations on quantum devices}, 
  author={Norm M. Tubman and Carlos Mejuto-Zaera and Jeffrey M. Epstein and Diptarka Hait and Daniel S. Levine and William Huggins and Zhang Jiang and Jarrod R. McClean and Ryan Babbush and Martin Head-Gordon and K. Birgitta Whaley},
  year={2018},
  eprint={1809.05523},
  archivePrefix={arXiv},
  primaryClass={quant-ph},
  url={https://arxiv.org/abs/1809.05523}, 
}

\section*{Figure Legends}
\paragraph{Figure~\ref{fig: problem_fig}:} \textbf{Variational quantum algorithm with guiding state.}
(a) The algorithm starts with a guiding state as a warm start $\rho_0$ and goes through a parameterized quantum circuit $U(\boldsymbol{\theta})$ to learn the representation of the ground state that will be used to generate the properties of the system. The new parameters are updated using gradient descent with learning rate $\eta$ with respect to the loss function $L(\boldsymbol{\theta})$.(b) We present our perspective on warm starts, which is similar to the conventional approach~\cite{drudis2024variational}. However, our focus shifts from the initialization of parameters to the initialization of the quantum state itself. It is worth noting that these two approaches can be mapped onto one another.

\paragraph{Figure~\ref{fig: ala}: \textbf{Alternating Layer Ansatz}.} (a) An illustration of the alternating layered ansatz. Here, each layer is separated by a vertical dashed line. (b) We illustrate the locality property of ALA. The shaded boxes are in the light cone of $O_i$. This means that the actions of $U^{\dagger}(\boldsymbol{\theta})O_iU(\boldsymbol{\theta})$ only depend on the parameters in the shaded boxes and the other will be canceled out.

\paragraph{Figure~\ref{fig:num_exp}: \textbf{Predicting ground state properties in 2D antiferromagnetic random Heisenberg models.}} (a) The variance of a single entry $K_{\boldsymbol{\theta}(0)}(x, x\prime)$ over $100$ different initialization in a variety of system sizes $n$. The experiment corresponds to the ALA with $L=1$ and $m=2$. (b) The distribution of {$\Delta_t K_{\boldsymbol\theta}(x,x') := K_{\boldsymbol \theta(t+1)}(x,x')-K_{\boldsymbol\theta(t)}(x,x')$} across different system size settings $n$ over a range of $t$ from $1$ to $100$. As the system size goes to $20$, the values $K_{\boldsymbol{\theta}}(x, x\prime)$ asymptotically stays constant. (c) Training behavior of the true model corresponding to linearized model with initialized kernel $K_{\boldsymbol{\theta}(0)}$. The models are performed with ALA$(n=20,m=4,L=2)$. (d) The generalization error with three different training dataset sizes of $30,50,80$.

\paragraph{Figure~\ref{fig:num_exp2}: \textbf{Neural Tangent Kernel matrix $K_{\boldsymbol{\theta}(0)}$.}} (a) The kernel entry $K_{\boldsymbol{\theta}(0)}(x, x')$ is shown for various system sizes, calculated over 100 random parameter initializations, with red crosses indicating the mean values. (b) The kernel matrix value with ALA$(n=20,m=4,L=2)$ and $M=80$. Additionally, the largest eigenvalue of the kernel is approximately $5.5$.

\clearpage
\setcounter{lmma}{0}
\setcounter{thrm}{0}
\setcounter{equation}{0}
\renewcommand\appendixname{Supplementary Information}
\renewcommand{\thethrm}{SI.\arabic{thrm}}
\renewcommand{\thelmma}{SI.\arabic{lmma}}
\renewcommand{\theequation}{S\arabic{equation}}
\appendix
\renewcommand{\thesection}{\arabic{section}}
\begin{center}
    \Large{\textbf{Supplemental Information for `Theoretical Guarantees of Variational Quantum Algorithm with Guiding States'}}
\end{center}

This Supplementary Information provides additional technical details and complete proofs of the main theoretical results presented in the manuscript. In Supplementary Notes~1 and~2, we establish the concentration properties of the quantum neural tangent kernel by proving Theorems~\ref{thrm: concentration_paper_version} and~\ref{thrm: lazy_paper_version}, respectively. Meanwhile, we give the full proofs of the convergence and generalization guarantees in Theorems~\ref{thrm: convergence_paper_ver} and~\ref{thrm: main_theorem} in Supplementary Notes~3 and~4.

\section*{Supplementary Note 1: Proof of Theorem~\ref{thrm: concentration_paper_version}} \label{si1}

Before proceeding to the detailed proofs, we first present the proof of Lemma~\ref{lmma: locality}. For convenience, we restate the lemma below.
\begin{lmma}
    \label{lmma: locality_app} Let a parameterized quantum circuit $U(\boldsymbol{\theta}) \in \text{ALA}(n,m,p, L)$. For any $k$-local observable $O$ such that $m \geq k$, $U^{\dagger}(\boldsymbol{\theta})OU(\boldsymbol{\theta})$ acts non-trivially on $\mathcal{O}(Lm)$ qubits.
\end{lmma}
\begin{proof}
    Let $S_{V}$ as the support of operator $V$. Denote $W_{L}$ is the block at the last layer such that $S_O \subseteq S_{W_L}$. 
    We are interested in finding the locality of $O(\boldsymbol{\theta}) = U^{\dagger}(\boldsymbol{\theta})OU(\boldsymbol{\theta})$. Noting that, for each layer $S_{O(\boldsymbol{\theta})}$ is extended by $2m$ qubits. So the locality of $O(\boldsymbol{\theta})$ is $\mathcal{O}(Lm)$. 
\end{proof}
Now, we provide a proof for Theorem~\ref{thrm: concentration_paper_version}, which we restated more details here for convenience.
\begin{thrm}[Concentration of initialization]
    Consider the dataset $\mathcal{S} = \{(x_i, y_i)\}_{i=1}^{M}$ and loss function $L_{\mathcal{S}}(\boldsymbol{\theta}) = \frac{1}{2M}\sum_{i=1}^{M} (f_{\boldsymbol{\theta}}(x_i) - y_i)^2$ with the model function defined on $U(\boldsymbol{\theta}) \in \text{ALA}(n,m,p,L)$. {Suppose that the parameters $\boldsymbol{\theta} \in \mathbb{R}^{L \times p \times n/m}$ are drawn from a product distribution; equivalently, the entries of $\boldsymbol{\theta}$ are initialized independently}. Then, the entry in the neural tangent kernel satisfies:
     \begin{align}
        \mathbb{P}_{\boldsymbol{\theta}}[\left|K_{\boldsymbol{\theta}}(x, x') - \mathbb{E}_{\boldsymbol{\theta}}[K_{\boldsymbol{\theta}}(x, x')]\right| \geq \epsilon] &\leq \exp{-\Omega(\frac{M^2K^4\epsilon^2}{L^3pnmc^4})}
    \end{align}
    for all $x, x' \in \mathcal{S}$.
    \label{thrm: concentration}
\end{thrm}
\begin{proof}
    The proof adapts from Theorem 1 \cite{abedi2023quantum}. Consider the model function:
    \begin{equation}
        f_{\boldsymbol{\theta}}(x) = \frac{1}{{K}} \sum_{\ell=1}^{K}\Tr[\rho_0(x) U^{\dagger}(\boldsymbol{\theta})O_\ell U(\boldsymbol{\theta})] = \frac{1}{{K}} \sum_{\ell=1}^{K}f^{\ell}_{\boldsymbol{\theta}}(x)
    \end{equation}
    The tangent kernel can be represented as:
    \begin{align}
        K_{\boldsymbol{\theta}}(x, x') &= \frac{1}{M}\nabla_{\boldsymbol{\theta}}f(\boldsymbol{\theta}, x)\cdot\nabla_{\boldsymbol{\theta}}f(\boldsymbol{\theta}, x') \\
                                       &= \frac{1}{MK^2} \sum_{\ell, \ell'=1}^{K}\sum_{j=1}^{L \times p\times n/m} \frac{\partial f^\ell_{\boldsymbol{\theta}}(x)}{\partial \theta_j}\cdot\frac{\partial f^{\ell'}_{\boldsymbol{\theta}}(x')}{\partial \theta_j}
    \end{align}
    Define $\mathcal{N}_\ell$ is the set of indices of parameters which $f^\ell_{\boldsymbol{\theta}}(x)$ depends on. Since $O_\ell$ is a k-local observable, following Lemma~\ref{lmma: locality_app}, $f^\ell_{\boldsymbol{\theta}}(x)$ acts non-trivially on $\mathcal{O}(Lm)$ qubits. On the other hand, each block contains $p$ parameters. Thus, $|\mathcal{N}_\ell| = \mathcal{O}(Lmp)$.
    
    \begin{align}
        K_{\boldsymbol{\theta}}(x, x') &= \frac{1}{MK^2} \sum_{\ell, \ell'=1}^{K}\sum_{j=\mathcal{N}_\ell \cap \mathcal{N}_{\ell'}} \frac{\partial f^\ell_{\boldsymbol{\theta}}(x)}{\partial \theta_j}\cdot\frac{\partial f^{\ell'}_{\boldsymbol{\theta}}(x')}{\partial \theta_j}
    \end{align}
    The equality holds since if $j \notin \mathcal{N}_\ell$, then $\frac{\partial f^\ell_{\boldsymbol{\theta}}(.)}{\partial \theta_j} = 0$.
    
    Define $\Gamma = \{(j, \ell, \ell'): j \in \mathcal{N}_\ell \cap \mathcal{N}_{\ell'} \forall \ell, \ell'\}$ and $T_{\ell, \ell', j}(\boldsymbol{\theta}) = \frac{\partial f^\ell_{\boldsymbol{\theta}}(x)}{\partial \theta_j}\cdot\frac{\partial f^{\ell'}_{\boldsymbol{\theta}}(x')}{\partial \theta_j}$, then:
    \begin{equation}
         K_{\boldsymbol{\theta}}(x, x') = \frac{1}{MK^2}\sum_{\ell, \ell'=1}^{K} \sum_{(j, \ell, \ell')\in \Gamma} T_{\ell, \ell', j}(\boldsymbol{\theta})
    \end{equation}
    The concentration bound is obtained by McDiarmid's inequality.
    Let $\boldsymbol{\theta}$ and $\boldsymbol{\theta'}$ $\in \mathbb{R}^{L\times p\times n/m}$ differ by only $j$-th entry. 
    \begin{align}
        \left| K_{\boldsymbol{\theta}}(x, x') -  K_{\boldsymbol{\theta'}}(x, x')\right| &\leq \frac{1}{MK^2} \sum_{\ell, \ell': (j, \ell, \ell')\in \Gamma} \left| T_{\ell, \ell', j}(\boldsymbol{\theta}) - T_{\ell, \ell', j}(\boldsymbol{\theta'})\right| \\
        &\leq \frac{1}{MK^2} \sum_{\ell, \ell': (j, \ell, \ell')\in \Gamma} \left| T_{\ell, \ell', j}(\boldsymbol{\theta})\right| + \left|T_{\ell, \ell', j}(\boldsymbol{\theta'})\right| \\
        &\in \mathcal{O}\left( \frac{c^2 m}{MK^2}\right)
    \end{align}
    Here, the last inequality comes from our assumption of~\eqref{eqn: assumption} and the fact that if we fix $j$, the number of triples $(j, \ell, \ell')$ is in $\mathcal{O}(Lm)$. Using McDiarmid's inequality, we have:
    \begin{align}
        \mathbb{P}[\left|K_{\boldsymbol{\theta}}(x, x') - \mathbb{E}[K_{\boldsymbol{\theta}}(x, x')]\right| \geq \epsilon] &\leq 2\exp{-\frac{2\epsilon^2}{Lpn/m\mathcal{O}(c^4L^2m^2/M^2K^4)}} \\
        &= \exp{-\Omega(\frac{M^2K^4\epsilon^2}{L^3pnmc^4})}
    \end{align}
\end{proof}
Since $L, m, p \in \mathcal{O}(\log(n))$ and $K = \mathrm{poly}(n)$, we obtain the result in Theorem~\ref{thrm: concentration_paper_version}. 

\section*{Supplementary Note 2: Proof of Theorem~\ref{thrm: lazy_paper_version}}\label{si2}
Theorem~\ref{thrm: concentration} implies that for every initialization of parameters, $\boldsymbol{\theta}$, the tangent kernel vanishes exponentially towards its mean. Furthermore, this kernel indeed stays constant along the gradient descent updates at the limit of $n$ goes to infinity. Particularly, we show that with the input of a guiding state~\eqref{eqn: guided_state}, if $\boldsymbol{\theta}$ is initialized from $\mathcal{N}(\boldsymbol{0}, \kappa^2\textbf{I})$, the tangent kernel $K_{\boldsymbol{\theta}}$ remains almost unchanged during the gradient descent algorithm when the system size is large enough.

We established the Theorem~\ref{thrm: lazy_paper_version} by combining the following Theorems~\ref{thrm: lazy} and~\ref{lmma: initial_loss}.
\begin{thrm}[Lazy Training]{Suppose that the initialized parameters $\boldsymbol{\theta}(0)$ are drawn from a product distribution.} Then, during the discrete gradient descent algorithm via loss function~\eqref{eqn: loss_function}, the change in any single entry $K_{ij}(\boldsymbol{\theta})$ of the tangent kernel $K_{\boldsymbol{\theta}}$~\eqref{eqn: ntk} at each discrete iteration $t$ is bounded by:
{\begin{equation}
    |K_{ij}(\boldsymbol{\theta}(t+1)) - K_{ij}(\boldsymbol{\theta}(t))| \leq \Tilde{\mathcal{O}}\left(\frac{\eta\cdot n}{M \cdot K^3}\sqrt{{L_{\mathcal{S}}(\boldsymbol{\theta}(0))}}\right)\ \ \  \forall i,j
\end{equation}}
\label{thrm: lazy}
\end{thrm}
\begin{proof}
{
Now, we can go into the proof of our main result. For convenience, we denote $\xi$ as the total number of parameters as such $\xi=n/m.L.p$. We wish to compute the finite difference $|K_{ij}(\boldsymbol{\theta}(t+1)) - K_{ij}(\boldsymbol{\theta}(t))|$ after one discrete gradient descent step. By the Mean Value Theorem, there exists a parameter vector $\tilde{\boldsymbol{\theta}}$ on the line segment connecting $\boldsymbol{\theta}(t)$ and $\boldsymbol{\theta}(t+1)$ such that:
\begin{align}
|K_{ij}(\boldsymbol{\theta}(t+1)) - K_{ij}(\boldsymbol{\theta}(t))| &\leq \sum_{u=1}^{\xi} \left| \partial_{\theta_u} K_{ij}(\tilde{\boldsymbol{\theta}}) \right|\cdot |\theta_u(t+1) - \theta_u(t)| \\
&= \sum_{u=1}^{\xi} \left| \frac{1}{MK^2}\sum_{(v, \ell, \ell')\in \Gamma} \partial_u\left(\partial_v f^\ell_{\tilde{\boldsymbol{\theta}}}(x_i) \partial_v f^{\ell'}_{\tilde{\boldsymbol{\theta}}}(x_j) \right)\right|\cdot |\theta_u(t+1) - \theta_u(t)|\\
&= \sum_{u=1}^{\xi} \left|\frac{1}{MK^2} \sum_{(v, \ell, \ell')\in \Gamma \text{ s.t. } u \in \mathcal{N}\ell \cup \mathcal{N}{\ell'}} \partial_u\left(\partial_v f^\ell_{\tilde{\boldsymbol{\theta}}}(x_i) \partial_v f^{\ell'}_{\tilde{\boldsymbol{\theta}}}(x_j) \right)\right|\cdot |\theta_u(t+1) - \theta_u(t)| \label{eqn: temp hi}\\
&\leq \sum_{u=1}^{\xi} \left|\frac{1}{MK^2} \sum_{(v, \ell, \ell')\in \Gamma \text{ s.t. } u \in \mathcal{N}\ell \cup \mathcal{N}{\ell'}} c^2 \right| \cdot |\theta_u(t+1) - \theta_u(t)| \label{eqn: temp}
\end{align}
Here, we use our assumption of~\eqref{eqn: assumption}. The transition from~\eqref{eqn: temp hi} to~\eqref{eqn: temp} holds if the partial differential operators $\partial_{\theta_u}, \partial_{\theta_v}$ either commute or anti-commute. This condition is generally true when generators of $U(\boldsymbol{\theta})$ are Pauli operators.On the other hand, the discrete parameter step magnitude is given by the gradient descent rule:
\begin{align}
    |\theta_u(t+1) - \theta_u(t)| &= \left| -\eta \frac{\partial L_{\mathcal{S}}(\boldsymbol{\theta}(t))}{\partial {\theta_u}}\right| \\
    &= \eta\left|\partial_{\theta_u} \left(\frac{1}{2M}\sum_{i=1}^{M} \left(f_{\boldsymbol{\theta}(t)}(x_i)-y_i\right)^2 \right)\right|\\
    &= \eta\left|\frac{1}{M}\sum_{i=1}^{M} \left(f_{\boldsymbol{\theta}(t)}(x_i)-y_i\right). \partial_{\theta_u} f_{\boldsymbol{\theta}(t)}(x_i)\right| \\
    &= \eta\left|\frac{1}{M}\sum_{i=1}^{M} \left(f_{\boldsymbol{\theta}(t)}(x_i)-y_i\right). \frac{1}{{K}}\sum_{\ell: u\in \mathcal{N}_\ell}\partial_{\theta_u} f^\ell_{\boldsymbol{\theta}(t)}(x_i)\right| \\
    &\leq \mathcal{O}\left( \eta\sqrt{\frac{L_{\mathcal{S}}(\boldsymbol{\theta}(t))}{K^2}}\right) \label{eqn: bound_grad_param}\\
    &\leq \mathcal{O}\left( \eta\sqrt{\frac{L_{\mathcal{S}}(\boldsymbol{\theta}(0))}{K^2}}\right) \label{eqn: loss_t_and_loss_0}
\end{align}
where~\eqref{eqn: bound_grad_param} uses our assumption~\eqref{eqn: assumption} and~\eqref{eqn: loss_t_and_loss_0} comes from the fact that gradient descent monotonically decreases the loss, ensuring $L_{\mathcal{S}}(\boldsymbol{\theta}(t)) \leq L_{\mathcal{S}}(\boldsymbol{\theta}(0))$. Substitute this bound back into~\eqref{eqn: temp} and use the fact that if we fix $(\ell, \ell')$, the number of elements in $\mathcal{N}_\ell \cup \mathcal{N}_{\ell'}$ is in $\mathcal{O}(Lmp)$ and $|\Gamma| = \mathcal{O}(\xi Lmp) = \Tilde{\mathcal{O}}(n)$ (as $L, m, p \in \mathcal{O}(\mathrm{log}(n))$), we have:
\begin{align}
    |K_{ij}(\boldsymbol{\theta}(t+1)) - K_{ij}(\boldsymbol{\theta}(t))| &\leq \left|\frac{1}{MK^2} c^2 \cdot \Tilde{\mathcal{O}}(n)\cdot \mathcal{O}\left( \eta\sqrt{\frac{L_{\mathcal{S}}(\boldsymbol{\theta}(0))}{K^2}}\right)\right|\\
    &\leq \Tilde{\mathcal{O}}\left(\frac{\eta\cdot n}{M\cdot K^3}\sqrt{{L_{\mathcal{S}}(\boldsymbol{\theta}(0))}}\right) 
\end{align}
}
\end{proof} 
The bounds of this theorem are effective when the loss function $L_{\mathcal{S}}(\boldsymbol{\theta})$ at initialization is a constant independent of $n$. The following theorem will specify the conditions under which this holds. 
\begin{thrm}
     Suppose $\boldsymbol{\theta}(0) \sim \mathcal{N}(\boldsymbol{0}, \kappa^2\mathbf{I})$ and $\kappa \in \Tilde{\mathcal{O}}(\frac{\sqrt{\gamma}\delta}{n})$, then, with probability at least $1 - \gamma$ over the random initialization 
     \begin{equation}
        L_{\mathcal{S}}(\boldsymbol{\theta}(0)) \leq \widetilde{\mathcal{O}}(\delta^2)
    \end{equation}
     \label{lmma: initial_loss}
\end{thrm}
\begin{proof}
    {
     First, each $\theta_i(0)$ has zero mean and variance of $\mathcal{O}(\kappa^2)$, which means $\mathbb{E}[(\theta_i(0))^2] = \mathcal{O}(\kappa^2)$. This implies $\mathbb{E}[\Vert \boldsymbol{\theta}(0) \Vert^2_2] = \mathcal{O}(\xi\kappa^2)$, where $\xi$ is the total number of parameters as such $\xi=n/m.L.p$ , and by Markov’s inequality we have $\Vert \boldsymbol{\theta}(0) \Vert^2_2 \leq \mathcal{O}(\xi\kappa^2/\gamma)$  with probability at least $1-\gamma$.} 
    {
    By Taylor's theorem with second-order remainder, there exists $\widetilde{\boldsymbol{\theta}}$ on the line segment joining $\boldsymbol{0}$ and $\boldsymbol{\theta}(0)$ such that
    \begin{align}
    f_{\boldsymbol{\theta}(0)}(x)
    &=
    f_{\boldsymbol 0}(x)
    +
    \nabla_{\boldsymbol\theta}
    f_{\boldsymbol\theta}(x)\big|_{\boldsymbol 0}^{\!\top}
    \boldsymbol{\theta}(0)
    +
    \frac12
    \boldsymbol{\theta}(0)^\top
    \nabla_{\boldsymbol\theta}^2
    f_{\boldsymbol\theta}(x)\big|_{\widetilde{\boldsymbol\theta}}
    \boldsymbol{\theta}(0) \\
    |f_{\boldsymbol{\theta}(0)}(x)-f_{\boldsymbol0}(x)| &\le \sqrt{\xi}c \|\boldsymbol{\theta}(0)\|_2 + \frac{\xi c}{2}\|\boldsymbol{\theta}(0)\|_2^2. \label{taylor_final}
    \end{align}
    where the last inequality use the smoothness assumption~\eqref{eqn: assumption}: $ \left\| \nabla_{\boldsymbol\theta} f_{\boldsymbol\theta}(x)\big|_{\boldsymbol0} \right\|_2 \le \sqrt{\xi}\,c$, and $ \left\| \nabla_{\boldsymbol\theta}^2 f_{\boldsymbol\theta}(x)\right\|_{2}\le \xi c$.
    Let us first note that without loss of generality, any block $W_{kl}(\boldsymbol{\theta}_{kl})$ in the ALA can be written as:
    \begin{equation}
        W_{kl}(\boldsymbol{\theta}_{kl}) = G_{p}(\theta^{p}_{kl})...G_{\nu}(\theta^{\nu}_{kl})...G_{1}(\theta^{1}_{kl})
    \end{equation}
    where each $G_{\nu}(\theta^{\nu}_{kl}) = e^{-i\theta_{kl}H_\nu}$, $H_{\nu}$ is a Hamiltonian describing the ALA. Thus, if $\boldsymbol{\theta} = \boldsymbol{0}$, then $U(\boldsymbol{\theta})$ consists a product of identity gates as such $U(\boldsymbol{0}) = \boldsymbol{I}$. We are interested to calculate $f_{\boldsymbol{0}}(x) - y$, where $y = \Tr[O\rho(x)]$:
    \begin{align}
        \left|f_{\boldsymbol{0}}(x) - y\right| &= \left|\Tr[O\rho_0(x)] - \Tr[O\rho^{*}(x)]\right| \\
        &\leq \Vert O \Vert_{\text{op}} \cdot 2\delta 
        \label{temp12}
    \end{align}
    where the last inequality comes from~\eqref{eqn: guided_state}. Combine~\eqref{temp12} to~\eqref{taylor_final}, we have:
    \begin{align}
        \left| f_{\boldsymbol{\theta}(0)}(x) -  f_{\boldsymbol{0}}(x) \right| &\leq \sqrt{\xi} c. \left\Vert \boldsymbol{\theta}(0) \right\Vert_2 {+ \frac{\xi c}{2}\|\boldsymbol{\theta}(0)\|_2^2} \\
        \left| (f_{\boldsymbol{\theta}(0)}(x) - y) -  (f_{\boldsymbol{0}}(x)-y)\right| &\leq \sqrt{\xi}c \left\Vert \boldsymbol{\theta}(0) \right\Vert_2 {+ \frac{\xi c}{2}\|\boldsymbol{\theta}(0)\|_2^2} \\
        \left|(f_{\boldsymbol{\theta}(0)}(x) - y)\right| -  \left|(f_{\boldsymbol{0}}(x)-y)\right| &\leq \sqrt{\xi}c \left\Vert \boldsymbol{\theta}(0) \right\Vert_2 {+ \frac{\xi c}{2}\|\boldsymbol{\theta}(0)\|_2^2} \\ 
        \left|(f_{\boldsymbol{\theta}(0)}(x) - y)\right| &\leq \Vert O \Vert_{\text{op}}.2\delta +  \sqrt{\xi}c.\left\Vert \boldsymbol{\theta}(0) \right\Vert_2 
        {+ \frac{\xi c}{2}\|\boldsymbol{\theta}(0)\|_2^2}
    \end{align}
    As a result, 
    \begin{align}
        L_{\mathcal{S}}(\boldsymbol{\theta}(0)) &\leq \frac{1}{2M}\sum_{i=1}^{M} \left(\Vert O \Vert_{\text{op}}.2\delta + \sqrt{\xi}c.\left\Vert \boldsymbol{\theta}(0) \right\Vert_2 {+ \frac{\xi c}{2}\|\boldsymbol{\theta}(0)\|_2^2} \right)^2 \\
        &\leq \mathcal{O}((2\delta + \xi\kappa/\sqrt{\gamma} {+ \xi^2\kappa^2/\gamma} )^2)
        \label{eqn: initial_loss}
    \end{align}
    From our choice $\kappa \in \Tilde{\mathcal{O}}(\frac{\sqrt{\gamma}\delta}{n})$ and $\xi = n/m.p.L$, where $m, L, p$ are in $\mathcal{O}(\log(n))$, we obtain $ \frac{\xi\kappa}{\sqrt{\gamma}} = \widetilde{\mathcal O}(\delta)$, and $\frac{\xi^2\kappa^2}{\gamma} = \widetilde{\mathcal O}\!\left({\delta^2}\right)$. Thus,
    \begin{align}
        L_{\mathcal{S}}(\boldsymbol{\theta}(0)) &\leq \widetilde{\mathcal{O}}(\delta^2)
        \label{eqn: initial_loss}
    \end{align}
    We complete the proof.
}

\end{proof}


\section*{Supplementary Note 3: Proof of Theorem~\ref{thrm: convergence_paper_ver}}\label{si3}


From the above results, we show that, at the limit $n \rightarrow \infty$, the tangent kernel $K_{\boldsymbol{\theta}}$ stays constant at initialization. That means the learning behavior of the model enters the lazy regime, so we could employ rigorous results from classical NTK theory to analyze the performance of our variational quantum algorithms. One immediate result we could derive from that is the linear model at the initialized kernel exhibits a linear convergence rate. In particular, we rewrite the updated rule of the linear model as follows:
\begin{equation}
    \hat{f}_{\boldsymbol{\theta}(t+1)} - \hat{f}_{\boldsymbol{\theta}(t)} = -\eta K_{\boldsymbol{\theta}(0)} (\hat{f}_{\boldsymbol{\theta}(t)} - y)
    \label{eqn: constant_step_size_linear_model}
\end{equation}
For brevity, we omit the inputs in this analysis.

\begin{thrm}[Linear convergence rate]
    Suppose $ 0 < \lambda_{\min}:= \lambda_{\min}(K_{\boldsymbol{\theta}(0)}) \leq \lambda_j \leq \lambda_{\max} := \lambda_{\max}(K_{\boldsymbol{\theta}(0)}) < \infty$ and for $\eta \leq \frac{1}{\lambda_{\max}}$. Then, under the gradient descent algorithm, the linear model at the initialized kernel~\eqref{eqn: constant_step_size_linear_model} has the loss function at any time $t$ is
    \begin{equation}
        |\hat{L}_{\mathcal{S}}(\boldsymbol{\theta}(t))| \leq \sum_{j} (1-\eta \lambda_{j})^{2t}|L_{\mathcal{S}}(\boldsymbol{\theta}(0))| 
    \end{equation}
    \label{crl: crl_1}
\end{thrm}
\begin{proof}
    Consider the update rule of $\hat{f}_{\boldsymbol{\theta}(t+1)} - \hat{f}_{\boldsymbol{\theta}(t)} = -\eta K_{\boldsymbol{\theta}(0)} (\hat{f}_{\boldsymbol{\theta}(t)} - y)$. Recursively applying this rule from $0$ to $t$, we get:
    \begin{align}
        \hat{f}_{\boldsymbol{\theta}(t)} - y &= (\mathbf{I} - \eta K_{\boldsymbol{\theta}(0)})^t \cdot (f_{\boldsymbol{\theta}(0)} - y) \\
        \Vert \hat{f}_{\boldsymbol{\theta}(t)} - y \Vert_2 &\leq \Vert( \mathbf{I} - \eta K_{\boldsymbol{\theta}(0)})^t \Vert_F \cdot \Vert f_{\boldsymbol{\theta}(0)} - y\Vert_2 \\
        |\hat{L}_{\mathcal{S}}(\boldsymbol{\theta}(t))| &\leq \Vert (\mathbf{I} - \eta K_{\boldsymbol{\theta}(0)})^t \Vert_F^2 \cdot |L_{\mathcal{S}}(\boldsymbol{\theta}(0))|
    \end{align}
    Note that $\mathbf{I} - \eta K_{\boldsymbol{\theta}(0)}$ is positive semidefinite, because we have $\eta  \leq \frac{1}{\lambda_{\max}}$. This implies $\Vert( \mathbf{I} - \eta K_{\boldsymbol{\theta}(0)})^t \Vert^2_F = \sum_{j} (1-\eta \lambda_{j})^{2t}$. Thus, 
    \begin{align}
         |\hat{L}_{\mathcal{S}}(\boldsymbol{\theta}(t))| &\leq \sum_{j} (1-\eta \lambda_{j})^{2t}|L_{\mathcal{S}}(\boldsymbol{\theta}(0))| 
    \end{align}
\end{proof}
We will use this result to analyze the convergence of the true model. Before going to that, we first derive the deviation of the loss function of the true model from its linear version as in the following theorem.

\begin{thrm}
    Consider Algorithm~\ref{alg: main_alg}, under the gradient descent, let $\hat{L}_{\mathcal{S}}(\boldsymbol{\theta}(t))$ be the model training error of the asymptotic dynamics governed by the initialized kernel $K_{\boldsymbol{\theta}(0)}$ and $L_{\mathcal{S}}(\boldsymbol{\theta}(t))$ be the model training error of the true dynamics governed by the time-variance kernel $K_{\boldsymbol{\theta}(t)}$ starting with the same initialization $\boldsymbol{\theta}(0)$, we have:
    \begin{equation}
       \left| L_{\mathcal{S}}(\boldsymbol{\theta}(t)) - \hat{L}_{\mathcal{S}}(\boldsymbol{\theta}(t))\right| \leq \Tilde{\mathcal{O}}\left(\frac{n}{K^3}\eta^2 t^2 L_{\mathcal{S}}(\boldsymbol{\theta}(0))^{3/2}\right)
    \end{equation}
    where $L_{\mathcal{S}}(\boldsymbol{\theta}(0))$ is the training error of the two models at the same initialization.
    \label{thrm: training_loss_bound}
\end{thrm}
\begin{proof}
    {For convenience, we denote the functions $f(t,x)$ and $\hat{f}(t,x)$ as the model functions after the $t$-th discrete iteration of gradient descent under the dynamics of $K_{\boldsymbol{\theta}(t)}$ and $K_{\boldsymbol{\theta}(0)}$, respectively. And let $\boldsymbol{F}(t) = [f(t, x_1), \dots, f(t, x_M)]^{\top}$, $\hat{\boldsymbol{F}}(t) = [\hat{f}(t, x_1), \dots, \hat{f}(t, x_M)]^{\top}$, and $Y=[y_1,\dots, y_M]^\top$. Under the discrete update rule, we have:
    \begin{align}
        \boldsymbol{F}(t+1) &= \boldsymbol{F}(t) - \eta K_{\boldsymbol{\theta}(t)}.\left(\boldsymbol{F}(t)-Y \right) \\
        \hat{\boldsymbol{F}}(t+1) &= \hat{\boldsymbol{F}}(t) - \eta K_{\boldsymbol{\theta}(0)}.\left(\hat{\boldsymbol{F}}(t)-Y\right)
    \end{align}
    We wish to study the difference of the losses:
    \begin{align}
        \left| L_{\mathcal{S}}(\boldsymbol{\theta}(t)) - \hat{L}_{\mathcal{S}}(\boldsymbol{\theta}(t))\right| &= \left| \frac{1}{2M} \left( \left\Vert \boldsymbol{F}(t)-Y \right\Vert_2^2 - \left\Vert \hat{\boldsymbol{F}}(t)-Y \right\Vert_2^2\right) \right| \\
        &= \left| \frac{1}{2M} \left( \left\Vert \boldsymbol{F}(t)-Y \right\Vert_2 + \left\Vert \hat{\boldsymbol{F}}(t)-Y \right\Vert_2\right)\left( \left\Vert \boldsymbol{F}(t)-Y \right\Vert_2 - \left\Vert \hat{\boldsymbol{F}}(t)-Y \right\Vert_2\right) \right|\\
        &\leq \frac{1}{\sqrt{M}} \left| \sqrt{2L_{\mathcal{S}}(\boldsymbol{\theta}(0))}.\left( \left\Vert \boldsymbol{F}(t)-Y \right\Vert_2 - \left\Vert \hat{\boldsymbol{F}}(t)-Y \right\Vert_2\right)\right|\\
        &\leq \sqrt{\frac{2}{{M}}} \left| \sqrt{L_{\mathcal{S}}(\boldsymbol{\theta}(0))}. \left\Vert \boldsymbol{F}(t) - \hat{\boldsymbol{F}}(t) \right\Vert_2\right| \label{eqn: loss_diff}
    \end{align}
    where we use $L_{\mathcal{S}}(\boldsymbol{\theta}(t)) \leq L_{\mathcal{S}}(\boldsymbol{\theta}(0))$ and $\hat{L}_{\mathcal{S}}(\boldsymbol{\theta}(t)) \leq L_{\mathcal{S}}(\boldsymbol{\theta}(0))$ in the first inequality and the triangle inequality for the second one, and both models have the same initialization.
    Denote the difference vector $\Delta(t) = \boldsymbol{F}(t) - \hat{\boldsymbol{F}}(t)$. We analyze its evolution over discrete steps. Subtracting the update rules gives:
    \begin{align}
        \Delta(t+1) &= \Delta(t) - \eta K_{\boldsymbol{\theta}(t)}(\boldsymbol{F}(t)-Y) + \eta K_{\boldsymbol{\theta}(0)}(\hat{\boldsymbol{F}}(t)-Y) \\
        &= \Delta(t) - \eta (K_{\boldsymbol{\theta}(t)} - K_{\boldsymbol{\theta}(0)})(\boldsymbol{F}(t)-Y) - \eta K_{\boldsymbol{\theta}(0)}(\boldsymbol{F}(t) - \hat{\boldsymbol{F}}(t)) \\
        &= (I - \eta K_{\boldsymbol{\theta}(0)})\Delta(t) - \eta (K_{\boldsymbol{\theta}(t)} - K_{\boldsymbol{\theta}(0)})(\boldsymbol{F}(t)-Y)
    \end{align}
    Applying the triangle inequality, we obtain:
    \begin{align}
        \Vert \Delta(t+1) \Vert_2 &\leq \Vert I - \eta K_{\boldsymbol{\theta}(0)} \Vert_2 \Vert \Delta(t) \Vert_2 + \eta \Vert K_{\boldsymbol{\theta}(t)} - K_{\boldsymbol{\theta}(0)} \Vert_2 \Vert \boldsymbol{F}(t)-Y \Vert_2
    \end{align}
    Because $K_{\boldsymbol{\theta}(0)}$ is a positive semidefinite matrix and we choose $\eta \leq 1/\lambda_{\max}$, we have $\Vert I - \eta K_{\boldsymbol{\theta}(0)} \Vert_2 \leq 1$. Furthermore, since gradient descent monotonically decreases the loss, $\Vert \boldsymbol{F}(t)-Y \Vert_2 = \sqrt{2M L_{\mathcal{S}}(\boldsymbol{\theta}(t))} \leq \sqrt{2M L_{\mathcal{S}}(\boldsymbol{\theta}(0))}$. Thus:
    \begin{align}
        \Vert \Delta(t+1) \Vert_2 &\leq \Vert \Delta(t) \Vert_2 + \eta \sqrt{2M L_{\mathcal{S}}(\boldsymbol{\theta}(0))} \Vert K_{\boldsymbol{\theta}(t)} - K_{\boldsymbol{\theta}(0)} \Vert_2 \label{eqn: delta_recurrence}
    \end{align}
    From Theorem~\ref{thrm: lazy}, the change in the kernel matrix entries at each discrete step is bounded. Since $K(\cdot)$ is an $M \times M$ matrix, its spectral norm is bounded by $M$ times its maximum entry difference. Accumulating this over $\tau$ steps:
    \begin{align}
        \Vert K_{\boldsymbol{\theta}(\tau)} - K_{\boldsymbol{\theta}(0)} \Vert_2 &\leq \sum_{k=0}^{\tau-1} \Vert K_{\boldsymbol{\theta}(k+1)} - K_{\boldsymbol{\theta}(k)} \Vert_2 \\
        &\leq \tau \cdot M \cdot \Tilde{\mathcal{O}}\left(\frac{\eta \cdot n}{M \cdot K^2}\sqrt{\frac{L_{\mathcal{S}}(\boldsymbol{\theta}(0))}{K^2}}\right) \\
        &\leq \Tilde{\mathcal{O}}\left(\tau \eta \frac{n}{K^3} \sqrt{L_{\mathcal{S}}(\boldsymbol{\theta}(0))}\right)
    \end{align}
    Substituting this into our recurrence relation~\eqref{eqn: delta_recurrence} for an arbitrary step $\tau$:
    \begin{align}
        \Vert \Delta(\tau+1) \Vert_2 &\leq \Vert \Delta(\tau) \Vert_2 + \Tilde{\mathcal{O}}\left(\tau \eta^2 \frac{n \sqrt{M}}{K^3} L_{\mathcal{S}}(\boldsymbol{\theta}(0))\right)
    \end{align}
    Since both models start at the same initialization, $\Vert \Delta(0) \Vert_2 = 0$. Summing this recurrence from $\tau=0$ to $t-1$, we get:
    \begin{align}
        \Vert \Delta(t) \Vert_2 &\leq \sum_{\tau=0}^{t-1} \Tilde{\mathcal{O}}\left(\tau \eta^2 \frac{n \sqrt{M}}{K^3} L_{\mathcal{S}}(\boldsymbol{\theta}(0))\right) \\
        &\leq \Tilde{\mathcal{O}}\left(t^2 \eta^2 \frac{n \sqrt{M}}{K^3} L_{\mathcal{S}}(\boldsymbol{\theta}(0))\right) \label{eqn: delta_t}
    \end{align}
    Replace~\eqref{eqn: delta_t} to~\eqref{eqn: loss_diff}, we have:
    \begin{align}
        \left| L_{\mathcal{S}}(\boldsymbol{\theta}(t)) - \hat{L}_{\mathcal{S}}(\boldsymbol{\theta}(t))\right| &\leq \sqrt{\frac{2}{M}} \sqrt{L_{\mathcal{S}}(\boldsymbol{\theta}(0))} \cdot \Tilde{\mathcal{O}}\left(t^2 \eta^2 \frac{n \sqrt{M}}{K^3} L_{\mathcal{S}}(\boldsymbol{\theta}(0))\right) \\
        &\leq \Tilde{\mathcal{O}}\left(\frac{n}{K^3}\eta^2 t^2 L_{\mathcal{S}}(\boldsymbol{\theta}(0))^{3/2}\right)
    \end{align}
    }
\end{proof}
Combining Theorems~\ref{crl: crl_1} and~\ref{thrm: training_loss_bound}, we have the convergence of the true model (Theorem~\ref{thrm: convergence_paper_ver}) restated as follows.

\begin{thrm}
    Suppose $0 < \lambda_{\min}:= \lambda_{\min}(K_{\boldsymbol{\theta}(0)}) \leq \lambda_j \leq \lambda_{\max} := \lambda_{\max}(K_{\boldsymbol{\theta}(0)}) < \infty$ and for $\eta \leq \frac{1}{\lambda_{\max}}$, $\kappa \in \Tilde{\mathcal{O}}(\frac{\sqrt{\gamma}\delta}{n})$. Then, with probability at least $1 - \gamma$ over the random initialization, the model training error of the true dynamics governed by the time-variance kernel $K_{\boldsymbol{\theta}(t)}$ satisfies:
    \begin{equation}
        L_{\mathcal{S}}(\boldsymbol{\theta}(t)) \leq \Tilde{\mathcal{O}}\left(\sum_{j} (1-\eta \lambda_{j})^{2t}\delta^2+\frac{n}{K^3}\eta^2 t^2 \delta^3\right) \ \ \ \forall t>0
    \end{equation}
\end{thrm}
\begin{proof}
    From Theorem~\ref{thrm: training_loss_bound}, we have:
    \begin{align}
        L_{\mathcal{S}}(\boldsymbol{\theta}(t)) - \hat{L}_{\mathcal{S}}(\boldsymbol{\theta}(t)) &\leq \Tilde{\mathcal{O}}\left(\frac{n}{K^3}\eta^2 t^2 L_{\mathcal{S}}(\boldsymbol{\theta}(0))^{3/2}\right) \\ 
    \end{align}
    Here, we use $a-b \leq |a-b|\forall a,b \geq 0$. Then, applying Theorem~\ref{crl: crl_1} and Theorem~\ref{lmma: initial_loss}, we get:
    \begin{align}
        L_{\mathcal{S}}(\boldsymbol{\theta}(t)) &\leq \sum_{j} (1-\eta \lambda_{j})^{2t}|L_{\mathcal{S}}(\boldsymbol{\theta}(0))|  + \Tilde{\mathcal{O}}\left(\frac{n}{K^3}\eta^2 t^2 L_{\mathcal{S}}(\boldsymbol{\theta}(0))^{3/2}\right) \\ 
        &\leq\Tilde{\mathcal{O}}\left(\sum_{j} (1-\eta \lambda_{j})^{2t}\delta^2+\frac{n}{K^3}\eta^2 t^2 \delta^3\right)
    \end{align}
    This completes the proof.
\end{proof}
\section*{Supplementary Note 4: Proof of Theorem~\ref{thrm: main_theorem}}\label{si4}
Theorem~\ref{thrm: training_loss_bound} allows us to bound the training loss of the variational algorithm model via the asymptotic concentrated kernel at initialization. Next, we are interested in evaluating the generalization error bound of the model through Rademacher complexity.


Let us define $\mathcal{F}$ as the class of model functions that are obtained through a particular learning model. Consider the hypothesis space, the goal is to find some function in the hypothesis space that minimizes the expected error with respect to an unknown distribution $\mathcal{D}$. However, as we usually cannot directly access the distribution $\mathcal{D}$, we are rather interested in the empirical loss through training examples $\mathcal{S}$. The gap between the empirical and expected error is called the \textit{generalization error}, which determines the performance of the hypothesis function $f$ on the unseen data drawn from the unknown probability distribution:
\begin{equation}
    \text{gen}(\boldsymbol{\theta}) := \left| \mathbb{E}_{x\sim \mathcal{D}}[ \left| f_{\boldsymbol{\theta}}(x) - \Tr[O\rho(x)]\right|] - \mathbb{E}_{x\sim \mathcal{S}}[ \left| f_{\boldsymbol{\theta}}(x) - \Tr[O\rho(x)]\right|]\right|. \label{eqn: gen_err_app}
\end{equation}
The Rademacher complexity theory allows us to obtain the bounds of generalization error associated with learning from training data \cite{shalev2014understanding}. For convenience, we denote $\ell(f_{\boldsymbol{\theta}}(x)) = \left|f_{\boldsymbol{\theta}}(x) - \Tr[O\rho(x)]\right|$, the Rademacher complexity of a function space $\ell \circ \mathcal{F}$ with respect to training data $\mathcal{S}$ is defined as follows:
\begin{equation}
    R(\ell \circ \mathcal{F} \circ \mathcal{S}) := \frac{1}{M} \mathbb{E}_{\boldsymbol{\sigma} \in \{-1, 1\}^{M}} \left[\sup_{f_{\boldsymbol{\theta}} \in \mathcal F} \sum_{i=1}^{M} \sigma_i \ell(f_{\boldsymbol{\theta}}(x_i))\right]
\end{equation}
This quantity provides a bound of the generalization error by the following lemmas
\begin{lmma}[Theorem 26.5 \cite{shalev2014understanding}]
    For a training sample $\mathcal{S} = \{x_1,\dots, x_M\}$ generated by an unknown distribution $\mathcal{D}$ and real-value function class $\mathcal{F}$, such that for all $x$ and $f\in \mathcal{F}$ we have $|\ell(f(x))| \leq c$. Then, for a confidence parameter $\gamma \in (0,1)$, with probability at least $1-\gamma$ over the choice of $\mathcal{S}$, every $f\in \mathcal{F}$ satisfies:
    \begin{equation}
        \mathbb{E}_{x\sim\mathcal{D}}[\ell(f(x))] - \mathbb{E}_{x\sim\mathcal{S}}[\ell(f(x))] \leq 2 R(\ell \circ \mathcal{F} \circ \mathcal{S}) + 4c\sqrt{\frac{2\ln(4/\gamma)}{M}}
    \end{equation}
    \label{lmma: generalization_error}
\end{lmma}
Using Lemma~\ref{lmma: generalization_error} and standard Rademacher symmetry, we obtain the following two-side generalization bound.
\begin{lmma}
\label{lem:two-sided-rademacher}
{Let $\mathcal G$ be a class of functions
$g:\mathcal X\to[0,c]$, and let
$\mathcal S=\{x_1,\ldots,x_M\}$ be sampled independently from
$\mathcal D$. Then, for every
$\gamma \in(0,1)$, with probability at least
$1-\gamma$ over the draw of $\mathcal S$, every
$g\in\mathcal G$ satisfies
\begin{equation}
\left|
\mathbb E_{x\sim\mathcal D}[g(x)]
-
\mathbb E_{x\sim\mathcal S}[g(x)]
\right|
\leq
2R(\mathcal G\circ\mathcal S)
+
4c
\sqrt{
\frac{
2\ln(8/\gamma)
}{M}
}.
\label{eqn:two-sided-rademacher-bound}
\end{equation}}
\end{lmma}
\begin{proof}
{Apply Lemma~\ref{lmma: generalization_error} to $\mathcal G$ with failure probability
$\gamma/2$. This gives
\begin{equation}
\mathbb E_{\mathcal D}[g]
-
\mathbb E_{\mathcal S}[g]
\leq
2R(\mathcal G\circ\mathcal S)
+
4c
\sqrt{
\frac{
2\ln(8/\gamma)
}{M}
}.
\end{equation}
For the opposite tail, apply Lemma~\ref{lmma: generalization_error} to the complemented class
\begin{equation}
c-\mathcal G
:=
\{x\mapsto c-g(x):g\in\mathcal G\}.
\end{equation}
Since
\begin{equation}
\mathbb E_{\mathcal D}[c-g]
-
\mathbb E_{\mathcal S}[c-g]
=
-
\left(
\mathbb E_{\mathcal D}[g]
-
\mathbb E_{\mathcal S}[g]
\right)
\end{equation}
where $\mathbb{E}_{S}[g] = M^{-1}\sum _i g(x_i)$, by standard Rademacher calculus (Lemma 26.6~\cite{shalev2014understanding}), we have
\begin{equation}
R((c-\mathcal G)\circ\mathcal S)
\leq
R(\mathcal G\circ\mathcal S).
\end{equation}
Thus, the same bound holds for the negative of the generalization gap. A union bound proves the claim.}
\end{proof}
Apply Lemma~\ref{lem:two-sided-rademacher} with $\mathcal{G} = \ell \circ \mathcal{F}$, we obtain:
\begin{equation}
     \left|\mathbb{E}_{x\sim\mathcal{D}}[\ell(f(x))] - \mathbb{E}_{x\sim\mathcal{S}}[\ell(f(x))]\right| \leq 2 R(\ell \circ \mathcal{F} \circ \mathcal{S}) + 4c\sqrt{\frac{2\ln(8/\gamma)}{M}}
\end{equation}
This indicates that the generalization error~\eqref{eqn: gen_err_app} is upper-bounded by the Rademacher complexity. If the quantity $R(\ell \circ \mathcal{F} \circ \mathcal{S})$ is small, then the target function could be learned reliably. Thus, we next aim to bound this quantity. First, we consider the Rademacher complexity of the function class defined by the linear model~\eqref{eqn: linearized_model}. Then, we analyze the asymptotic result of the true function class generated by time-dependent kernel $K_{\boldsymbol{\theta}(t)}$.

{Let $\widehat{\mathcal F}$ denote the function space generated by the
linearized model. We define the signed residual function at time $t$ by
$
    r_t(x)
    :=
    \widehat f_{\boldsymbol{\theta}(t)}(x)-y(x),
$
where
$
y(x)=\Tr[O\rho(x)]
$
is the target function. On the training sample
$\mathcal S=\{x_i\}_{i=1}^M$, we collect these residuals into the vector $\mathbf r_t := (r_t(x_1), \dots,r_t(x_M))^{\top}$. The corresponding sample-wise absolute-loss function and loss vector are defined by $\ell_t(x):=|r_t(x)|$ and $\boldsymbol{\ell}_t:=|\mathbf r_t|$, respectively. Thus, the dynamics of the linearized model can be written in terms of the signed residual as
\begin{equation}
    \frac{d}{dt}r_t
    =
    -\eta\Pi r_t,
    \label{eqn: residual_gradient_flow}
\end{equation}
where the kernel operator $\Pi$ is defined by
\begin{equation}
    (\Pi g)(x)
    :=
    \sum_{i=1}^{M}
    g(x_i)
    K_{\boldsymbol{\theta}(0)}(x_i,x)
\end{equation}
for any function $g$ defined on the input space. Since $K_{\boldsymbol{\theta}(0)}$ is fixed with respect to the training time $t$,~\eqref{eqn: residual_gradient_flow} has the solution $r_t=e^{-\eta t\Pi}r_0$, with $r_0=f_{\boldsymbol{\theta}(0)}-y$. Consequently, the absolute-loss function satisfies
$\ell_t=|r_t|=\left|e^{-\eta t\Pi}r_0\right|$, where the absolute value is taken pointwise.}

{We decompose $r_0 = \Delta(\phi_0)+ \Delta(\phi_1)+\dots+ \Delta(\phi_M)$ along the eigenspace of $\Pi$, where $\Delta(\phi_0)$ is in the kernel (null-space) of $\Pi$ and $\Delta(\phi_i) \propto \phi_i$, then follows the discrete-time dynamics we obtain:
\begin{equation}
    r_t = \Delta(\phi_0) + \sum_{i=1}^{M}(1-\eta\lambda_i)^t\Delta(\phi_i)
\end{equation}
We see that the convergence of $r_t$ is faster along the eigenspaces with larger eigenvalues $\lambda_i$. We are typically interested in the case where the model focuses on fitting the most relevant kernel principal components (larger eigenvalues). For the subsequent Rademacher-complexity analysis, define the admissible
set of training times
\begin{equation}
    \mathcal T_B
    :=
    \left\{
    t\in\mathbb N
    \;\middle|\;
    B_1
    \leq
    \sum_{i=1}^{M}
    (1-\eta\lambda_i)^t
    \leq
    B_2
    \right\}.
    \label{eqn: admissible_training_times}
\end{equation}
As the absolute-loss function $\ell_t=|r_t|$, we later will use
the Rademacher contraction principle for the $1$-Lipschitz map
$z\mapsto |z|$. The corresponding class of absolute-loss functions generated by the
linearized trajectory is
\begin{equation}
    (\ell\circ\widehat{\mathcal F})_B
    :=
    \left\{
    x\mapsto |r_t(x)|
    \;\middle|\;
    t\in\mathcal T_B
    \right\}.
    \label{eqn: bounded_func_class}
\end{equation}
}

\begin{thrm}[Generalization error at initialization kernel]
    Consider a learning model trained the dataset $\mathcal{S} = \{x_1,\dots, x_M\}$ with learning rate $\eta \leq \frac{1}{\lambda_{\max}}$, which governed by a deterministic kernel $K_{\boldsymbol{\theta}(0)}$ at initialization, then the Rademacher complexity of the class $(\ell\circ\hat{\mathcal{F}})_B$ satisfied:
    \begin{equation}
        R((\ell\circ \hat{\mathcal{F}})_B\circ\mathcal{S}) \leq  B_2\sqrt{\frac{2}{M}L_{\mathcal{S}}(\boldsymbol{\theta}(0))}
    \end{equation}
    \label{thrm: linear_gen}
\end{thrm}

\begin{proof}
{Before proceeding, we recall the Rademacher contraction principle.
\begin{lmma}[Lemma 26.9~\cite{shalev2014understanding}]
\label{lem:rademacher-contraction}
{Let $\mathcal G$ be a class of real-valued functions on $\mathcal X$,
and let $\psi:\mathbb R\to\mathbb R$ be an $L$-Lipschitz function
satisfying $\psi(0)=0$. Then, for every sample
$\mathcal S=\{x_1,\ldots,x_M\}$,
\begin{equation}
R\bigl((\psi\circ\mathcal G)\circ\mathcal S\bigr)
\leq
L\cdot R\bigl(\mathcal G\circ\mathcal S\bigr).
\label{eq:rademacher-contraction}
\end{equation}}
\end{lmma}
Since the absolute-value map $\psi(z)=|z|$ is $1$-Lipschitz and satisfies $\psi(0)=0$, Lemma~\ref{lem:rademacher-contraction} gives
\begin{equation}
R\bigl(
(\ell\circ\widehat{\mathcal F})_B\circ\mathcal S
\bigr)
\leq
R\bigl(
(r\circ\widehat{\mathcal F})_B\circ\mathcal S
\bigr).
\label{eqn:loss-residual-contraction}
\end{equation}
}
{Given the kernel $K_{\boldsymbol{\theta}(0)}$ defined on the training samples $\mathcal{S}$, then for every $x\in \mathcal{S}$ we have:
    \begin{align}
        r_t(x) &= \sum_{i}^{M}r_0(x_i) \sum_{j}^{M} (1-\eta \lambda_j)^t {\phi_j(x_i)\cdot\phi_j(x)} \\
        &= \sum_{j=1}^{M} \left({ \sum_{i}^{M}r_0(x_i) (1-\eta \lambda_j)^t \phi_j(x_i)}\right)\cdot {\phi_j(x)}\\
        &=\sum_{j=1}^{M} {\omega_{j}(t)}\cdot{\phi_j(x)}
    \end{align}
    Here, we denote ${\omega_{j}(t)} = \sum_{i}^{M}r_0(x_i) (1-\eta \lambda_j)^t {\phi_j(x_i)}$. The Rademacher complexity of a function class is defined as:
    \begin{align}
        R((r\circ \hat{\mathcal{F}})_B\circ\mathcal{S}) &= \mathbb{E}_{\boldsymbol{\sigma}\sim \{-1,1\}^{M}} \left[\frac{1}{M}\sup_t \sum_{i=1}^{M} \sigma_i r_t(x_i)\right] =  \mathbb{E}_{\boldsymbol{\sigma}\sim \{-1,1\}^{M}} \left[\frac{1}{M}\sup_t \sum_{i=1}^{M} \sigma_i \left(\sum_{j=1}^{M}{\omega_{j}(t)}\cdot{\phi_j(x_i)}\right)\right]\\
        &= \mathbb{E}_{\boldsymbol{\sigma}\sim \{-1,1\}^{M}} \left[\frac{1}{M}\sup_t \sum_{i,j=1}^{M} \sigma_i \left({\omega_{j}(t)}\cdot{\phi_j(x_i)}\right)\right]\\
        &= \mathbb{E}_{\boldsymbol{\sigma}\sim \{-1,1\}^{M}} \left[\frac{1}{M}\sup_t \sum_{j=1}^{M} \left({\omega_{j}(t)}\cdot{\sum_{i=1}^{M} \sigma_i\phi_j(x_i)}\right)\right]\\
        &\leq \mathbb{E}_{\boldsymbol{\sigma}\sim \{-1,1\}^{M}} \left[\frac{1}{M}\sup_t \sum_{j=1}^{M} \Vert\omega_{j}(t)\Vert. \left\Vert{\sum_{i=1}^{M} \sigma_i\phi_j(x_i)}\right\Vert\right]\\
        &\leq \mathbb{E}_{\boldsymbol{\sigma}\sim \{-1,1\}^{M}} \left[\frac{1}{M}\sup_t \sum_{j=1}^{M} \Vert\omega_{j}(t)\Vert. \left(\sqrt{{\sum_{i=1}^{M} \sigma_i\phi_j(x_i)}\sum_{i'=1}^{M} \sigma_{i'}\phi_j(x_{i'})}\right)\right]\\
        &\leq \mathbb{E}_{\boldsymbol{\sigma}\sim \{-1,1\}^{M}} \left[\frac{1}{M}\sup_t \sum_{j=1}^{M} \Vert\omega_{j}(t)\Vert. \left(\sqrt{\sum_{i, i'=1}^{M} \sigma_i\sigma_{i'} {\phi_j(x_i)}\cdot{\phi_j(x_{i'})}}\right)\right]
    \end{align}
Let $t_-:=\min\mathcal T_B$, for every $t\in\mathcal T_B$, $|\omega_j(t)| \leq |\omega_j(t_-)|$ as $0\leq (1-\eta\lambda_j) < 1$. Hence,
\begin{align}
R((r\circ\hat{\mathcal F})_B\circ\mathcal S)
&\leq
\frac{1}{M}
\sum_{j=1}^{M}
\Vert\omega_j(t_-)\Vert
\mathbb E_{\boldsymbol{\sigma}}
\left[
\sqrt{
\sum_{i,i'=1}^{M}
\sigma_i\sigma_{i'}
\phi_j(x_i)\cdot\phi_j(x_{i'})
}
\right].
\label{eqn:rad_after_sup}
\end{align}
Here, we apply the property of eigenfunctions $\{\phi_j\}$: $\sum_{i=1}^{M}{\phi_j(x_i)}{\phi_j(x_i)} = 1$ and ${\phi_j(x_{i'})}{\phi_j(x_i)}={\phi_j(x_i)}{\phi_j(x_{i'})}\forall i, i'$. Note that $\mathbb{P}[\sigma_i = 1]=\mathbb{P}[\sigma_i = -1]=1/2$, we have:
    \begin{align}
        \mathbb{E}_{\boldsymbol{\sigma}\sim \{-1,1\}^{M}} \sqrt{\sum_{i, i'=1}^{M} \sigma_i\sigma_{i'} {\phi_j(x_i)}{\phi_j(x_{i'})}} \leq \sqrt{\mathbb{E}_{\boldsymbol{\sigma}\sim \{-1,1\}^{M}} \sum_{i, i'=1}^{M} \sigma_i\sigma_{i'} {\phi_j(x_i)}{\phi_j(x_{i'})}} = 1
    \end{align}
Thus, 
    \begin{align}
        R((r\circ \hat{\mathcal{F}})_B\circ\mathcal{S}) &\leq \frac{1}{M} \sum_{j=1}^{M} \Vert\omega_{j}(t_-)\Vert\\
        &= \frac{1}{M} \sum_{j=1}^{M} \left|(1-\eta \lambda_j)^{t_-}\right| \left\Vert \sum_{i}^{M}r_0(x_i) \phi_j(x_i) \right\Vert\\
        &\leq \frac{1}{M} \sum_{j=1}^{M} \left|(1-\eta \lambda_j)^{t_-}\right| \sqrt{\left|2M\cdot L_{\mathcal{S}}(\boldsymbol{\theta}(0))\right|\cdot \sum_{i}^{M}|\phi_j(x_i)|^2}\\
        &= \sqrt{\frac{2}{M}} \sum_{j=1}^{M} \left|(1-\eta \lambda_j)^{t_-}\right| \sqrt{L_{\mathcal{S}}(\boldsymbol{\theta}(0))}\\
        &\leq  B_2  \sqrt{\frac{2L_{\mathcal{S}}(\boldsymbol{\theta}(0))}{M}}
    \end{align}
    The last inequality holds since $t_-\in\mathcal T_B$, by the definition of
$\mathcal T_B$,
$
\sum_{j=1}^{M}
(1-\eta\lambda_j)^{t_-}
\leq B_2
$. This completes the proof.}
\end{proof}

Now, we focus on the generalization error of the true model governed by the time-varying kernel $K_{\boldsymbol{\theta}(t)}$. We perform the asymptotic study of this via the analysis of the deterministic initialization kernel $K_{\boldsymbol{\theta}(0)}$. We denote $\mathcal{F}$ as the function class generated from the true dynamic~\eqref{eqn: grad_f}. The generalization error is bounded as follows:
\begin{thrm}
    {Consider Algorithm~\ref{alg: main_alg}, under the gradient descent with learning rate $\eta \leq \frac{1}{\lambda_{\max}}$, the model is governed by the time-variance kernel $K_{\boldsymbol{\theta}(t)}$. Then for a confidence parameter $\gamma \in (0,1)$ and for every $t\in \mathcal{T}_B$ satisfying {$t=\mathcal{O}\!\left( \frac{\ln M}{\eta\lambda_{\min}} \right)$}, with a probability at least $1-\gamma$, we have the generation error~\eqref{eqn: gen_err_app} at the time $t$ as follows:
    \begin{equation}
    \text{gen}(\boldsymbol{\theta}(t)) \leq 2 B_2\sqrt{\frac{2}{M}L_{\mathcal{S}}(\boldsymbol{\theta}(0))} + 4c\sqrt{\frac{2\ln(8/\gamma)}{M}} +\Tilde{\mathcal{O}}\left(\frac{n}{K^3} \left(\frac{\ln(M/B_1)}{\lambda_{\min}}\right)^2 L_{\mathcal{S}}(\boldsymbol{\theta}(0))\right) 
    \end{equation}
    where $\lambda_{\min}$ is the smallest eigenvalues of the initialized tangent kernel $K_{\boldsymbol{\theta}(0)}$}.
    \label{thrm: gen_error_theorem}
\end{thrm}

\begin{proof}
{Apply Lemma~\ref{lem:two-sided-rademacher}, to $(\ell\circ\mathcal F)_B$ gives, with
probability at least $1-\gamma$ over the draw of
$\mathcal S$,
\begin{equation}
\operatorname{gen}(\boldsymbol{\theta}(t))
\leq
2R((\ell\circ\mathcal F)_B\circ\mathcal S) 
+
4c
\sqrt{
\frac{
2\ln(8/\gamma)
}{M}
}, \quad \forall t\in \mathcal{T}_B.
\label{eqn:si8_two_sided_bound}
\end{equation}
By the triangle inequality,
\begin{align}
R((\ell\circ\mathcal F)_B\circ\mathcal S)
\leq\;&
R((\ell\circ\widehat{\mathcal F})_B\circ\mathcal S)+
\left|
R((\ell\circ\mathcal F)_B\circ\mathcal S)
-
R((\ell\circ\widehat{\mathcal F})_B\circ\mathcal S)
\right|.
\label{eqn:si8_complexity_triangle}
\end{align}
Substituting~\eqref{eqn:si8_complexity_triangle} into~\eqref{eqn:si8_two_sided_bound} gives
\begin{equation}
\operatorname{gen}(\boldsymbol{\theta}(t))\leq2R((\ell\circ\widehat{\mathcal F})_B\circ\mathcal S)+
2\left|
R((\ell\circ{\mathcal F})_B\circ\mathcal S)
-
R((\ell\circ\widehat{\mathcal F})_B\circ\mathcal S)
\right|
+
4c
\sqrt{
\frac{
2\ln(8/\gamma)
}{M}
}.\label{eqn:si8_gen_decomposition}
\end{equation}
We next compare the Rademacher complexities of the nonlinear and
linearized absolute-loss trajectories. Using
$
\left|\sup_t a_t-\sup_t b_t\right|
\leq
\sup_t|a_t-b_t|
$, we obtain
\begin{align}
\left|
R((\ell\circ\mathcal{F})_B\circ \mathcal S)
-
R((\ell\circ\widehat{\mathcal{F}})_B\circ\mathcal S)
\right|
&\leq
\frac1M
\mathbb E_\sigma
\sup_{t\in\mathcal T_B}
\left|
\sum_{i=1}^M
\sigma_i
\left(
\ell_{t,i}-\widehat\ell_{t,i}
\right)
\right|
\\
&\leq
\frac1M
\mathbb E_\sigma\|\sigma\|_2
\sup_{t\in\mathcal T_B}
\left\|
\boldsymbol\ell_t-\widehat{\boldsymbol\ell}_t
\right\|_2
\\
&=
\frac1{\sqrt M}
\sup_{t\in\mathcal T_B}
\left\|
\boldsymbol\ell_t-\widehat{\boldsymbol\ell}_t
\right\|_2.
\end{align}
Here, we reemploy the notation from Theorem~\ref{thrm: training_loss_bound}: $\boldsymbol{F}(t) = [f(t, x_1), \dots, f(t, x_M)]^{\top}$, $\widehat{\boldsymbol{F}}(t) = [\hat{f}(t, x_1), \dots, \hat{f}(t, x_M)]^{\top}$, and $Y=[y_1,\dots, y_M]^\top$. Thus, for each training point, the reverse triangle inequality gives
\begin{equation}
\left|
|F_{i}(t)-Y_i|
-
|\widehat F_{i}(t)-Y_i|
\right|
\leq
|F_{i}(t)-\widehat F_{i}(t)|.
\end{equation}
It follows that
\begin{equation}
\left\|
\boldsymbol\ell_t-\widehat{\boldsymbol\ell}_t
\right\|_2
\leq
\left\|
\mathbf F(t)-\widehat{\mathbf F}(t)
\right\|_2.
\label{eq:reverse-triangle-vector}
\end{equation}
Therefore,
\begin{equation}
\left|
R((\ell\circ\mathcal{F})_B\circ \mathcal S)
-
R((\ell\circ\widehat{\mathcal{F}})_B\circ\mathcal S)
\right|
\leq
\frac1{\sqrt M}
\sup_{t\in\mathcal T_B}
\left\|
\mathbf F(t)-\widehat{\mathbf F}(t)
\right\|_2.
\label{eq:complexity-difference-prediction-vector}
\end{equation}
Using~\eqref{eqn: delta_t}
\begin{equation}
\left|
R((\ell\circ\mathcal{F})_B\circ \mathcal S)
-
R((\ell\circ\widehat{\mathcal{F}})_B\circ\mathcal S)
\right|
\leq
\widetilde O\left(
\frac{\eta^2n}{K^3}\cdot\sup_{t\in\mathcal T_B}t^2 \cdot
L_{\mathcal S}(\boldsymbol{\theta}(0))
\right).
\label{eqn:si8_complexity_difference_t}
\end{equation}
It remains to bound $t$ uniformly over $\mathcal T_B$. For every
$t\in\mathcal T_B$,
$
B_1
\leq
\sum_{j=1}^M
(1-\eta\lambda_j)^t \leq M(1-\eta\lambda_{\min})^t,
$
as $\lambda_j\geq\lambda_{\min}$ and $0\leq1-\eta\lambda_j\leq1$. Therefore,
\begin{equation}
t
\leq
\frac{
\ln(M/B_1)
}{
-\ln(1-\eta\lambda_{\min})} \leq
\frac{
\ln(M/B_1)
}{
\eta\lambda_{\min}
}.
\label{eqn:si8_time_bound}
\end{equation}
The last inequality uses $-\ln(1-z)\geq z, \text{for }0\leq z<1$.
Then we have implies $t=\mathcal{O}\left(
\frac{\ln M}
{\eta\lambda_{\min}}
\right)
$. In particular, the upper bound in~\eqref{eqn:si8_time_bound} implies
$
t^2
\leq
\left(
\frac{\ln(M/B_1)}
{\eta\lambda_{\min}}
\right)^2.
$
Substituting this inequality into~\eqref{eqn:si8_complexity_difference_t}, we obtain
\begin{align}
&
\left|
R((\ell\circ\mathcal F)_B\circ\mathcal S)
-
R((\ell\circ\widehat{\mathcal F})_B\circ\mathcal S)
\right|\leq
\widetilde{\mathcal O}\left[
\frac{n}{K^3}
\left(
\frac{\ln(M/B_1)}
{\lambda_{\min}}
\right)^2
L_{\mathcal S}(\boldsymbol{\theta}(0))
\right].
\label{eqn:si8_complexity_final}
\end{align}
Finally, we replace the result from Theorem~\ref{thrm: linear_gen} that $ R((\ell\circ \hat{\mathcal{F}})_B\circ\mathcal{S}) \leq  B_2\sqrt{\frac{2}{M}L_{\mathcal{S}}(\boldsymbol{\theta}(0))}$, together with~\eqref{eqn:si8_complexity_final} into~\eqref{eqn:si8_gen_decomposition}, we obtain
\begin{align}
    \text{gen}(\boldsymbol{\theta}(t)) \leq2B_2\sqrt{\frac{2}{M}L_{\mathcal{S}}(\boldsymbol{\theta}(0))} + 4c\sqrt{\frac{2\ln(8/\gamma)}{M}} +  \Tilde{\mathcal{O}}\left(\frac{n}{K^3} \left(\frac{\ln(M/B_1)}{\lambda_{\min}}\right)^2 L_{\mathcal{S}}(\boldsymbol{\theta}(0)) \right) \label{eqn: final}
\end{align}
}
\end{proof}
To obtain Theorem~\ref{thrm: main_theorem}, we apply Theorem~\ref{lmma: initial_loss} with failure probability $\gamma/2$ over the random initialization and Theorem~\ref{thrm: gen_error_theorem} with failure probability $\gamma/2$ over the draw of the training sample $\mathcal S$. By a union bound, both events hold simultaneously with probability at least $1-\gamma$ jointly over $\mathcal S$ and $\boldsymbol{\theta}(0)$. Finally, using Theorem~\ref{lmma: initial_loss}, we complete the proof.

\end{document}